# First Plasma Commissioning and Operational Highlights from India's First Spherical Tokamak at IPR

Kishore Mishra[a], Aditya Verma, N. Mansoori, Saurabh Verma, Y. Paravastu, M. S. Khan, Arvind Kumar, S. G. Thatipamula[a], Vishal Verma, M Sheetal, Mohit, U. Thaker, Ayush, Jignesh Patel, Praveenlal, Jagabandhu Kumar, F. S. Pathan, S. Ranjithkumar, S. Sam, Prasada Rao P., A. Jaiswal, S. Jha, Neelam Ramaiya, Utsav Rajvanshi, Santosh Pandya, K. Tahaliyani, S. Purohit, Suraj Gupta, Kiran Patel, Suman Aich, D. Kumawat, Deepak Kumar, Saurav Kumar, Rahul Kumar, Vismaysinh Raulji, Pramila, Praveena Kumari, Minsha Shah, Arun Prakash, Narendra Bhupesh, Amit K. Singh, V. Menon, Deepti Sharma, A. Kundu, G. K. Rajan, Prakash Parmar, Ankit Kumar, Vishnu Patel, Imran Mansuri, M Bhandarkar, H. Chudasma, Atish Sharma, Kirit Patel, H Masand, Prem Kumar, Parmesh Kumar, S Kannaujiya, Pramod Parmar, Arvind L. Thakur, C. G. Virani, Ankit Kumar, K. K. Ambulkar, Vilas Chaudhari, C. Danani, Y. S. S. Srinivas, S. Nair, K. Mahajan, R. Rajpal, M. B. Chowdhury, Manoj Kumar, S. K. Pathak, P. K. Sharma, Z. Khan, E. Rajendrakumar, J. Ghosh[a], and Team ST

*Institute for Plasma Research, Bhat, Gandhinagar 382 428*

[a]*Homi Bhabha National Institute, Training School Complex, Anushaktinagar, Mumbai 400 094*

E-mail : kmishra@ipr.res.in

**Abstract**

A compact Spherical Tokamak (ST) is commissioned at IPR to explore low aspect ratio tokamak physics and technologies that complement to the existing high aspect ratio tokamaks namely, ADITYA-U and SST-1 by enabling studies of non-inductive startup, current drive in over-dense plasmas, and shaped-plasma physics on a low-cost platform. The device, India's first spherical tokamak, has completed major mechanical, magnetic, and electrical integration, and the coil system has been successfully tested with series of integrated commissioning. First plasma experiments have been carried out with a modest Ohmic system assisted by a 2.45 GHz microwave system, supported by a centralized control and data acquisition system and an initial diagnostic set comprising imaging, spectroscopy, magnetic sensors, and radiation monitors. This paper presents the integrated commissioning experiences and first-plasma experiments of the newly installed machine.

**1. Introduction**

A spherical tokamak is a tokamak operated at low aspect ratio, typically with a tight central column and a plasma cross-section that nearly fills the torus [1,2]. The modern experimental basis for the ST concept was established through devices such as START, which demonstrated hot plasmas below aspect ratio 2 and confirmed several of the anticipated low-aspect-ratio advantages, including high beta, natural elongation, and vertical stability in appropriate operating regimes [3]. Later devices such as MAST, NSTX, PEGASUS, QUEST [4-9], and related programs expanded the ST physics base to include confinement, energetic particles, startup, divertor performance, and plasma control, thereby making the ST a credible route for both compact neutron-source and fusion-power concepts [10-15].

Fusion research through Spherical Tokamak concept became attractive with their ability to produce and sustain high beta plasma with natural elongation. The fusion power produced per volume scales with the square of plasma pressure (through beta) and with a high power of the toroidal field, so designing for large normalized beta $\beta_t$ and strong on-axis field $B_0$ is advantageous. Writing the plasma beta as $\beta = (2\mu_0 <n_eT_e>)/B^2$ , where $<n_eT_e>$ is volume averaged pressure and $B^2/2\mu_0$ is average magnetic pressure, shows the trade-off: for a given magnetic field the achievable pressure (and thus fusion source) grows with density and temperature, but the normalized beta $\beta_t = \beta/ (B_0^2)$ indicates how effectively the confining field is used. The stability limit expressed via $\beta_t \approx \beta_N I_p/(aB_0)$ shows that a high $\beta_N$ (normalized to engineering limits with $I_p$ as plasma current) and the ability to sustain large $I_p$ at modest $B_0$ directly improve the exploitable pressure before MHD limits are reached [16,17].

Because STs operate at low aspect ratio $A = R_0/a$ and naturally support strong shaping (high elongation κ and triangularity), the scaling for $I_p$ at fixed edge safety factor $q_a$ and for fusion gain both become favorable: plasma current and fusion gain scale positively with geometric shaping factors (elongation $\kappa$, $A$) and with $\beta_N$, while requiring less toroidal field for the same current or performance. In other words, the ST geometry concentrates the magnetic and current paths so that for a given $B_0$ and machine size an ST can access higher $\beta$ and bootstrap fractions, reducing reliance on external current drive and central-solenoid volt-seconds. These relations explain why STs are attractive as compact, high-performance devices: they offer a pathway to high fusion power density and efficient steady-state scenarios, provided center-stack engineering, plasma startup, and control challenges inherent to the low-$A$ topology are solved.

Although spherical tokamaks are increasingly being developed toward compact fusion reactor concepts, most major programs, including NSTX-U [18], MAST [19], and QUEST [20], have evolved through an initial phase of smaller experimental devices. At IPR, research on conventional high-aspect-ratio tokamaks has reached a considerable level of maturity over the years, whereas spherical tokamak plasma physics and engineering have not yet been explored. A laboratory-scale spherical tokamak therefore provides an important platform for investigating key issues that complement conventional tokamak research, such as non-inductive startup, topological plasma current drive, bootstrap current, electron Bernstein wave heating in over-dense plasmas, and high-harmonic fast-wave current drive. In this context, IPR has conceptualized and recently commissioned a small spherical tokamak to develop the required expertise in spherical tokamak engineering, operation, and associated plasma-physics challenges.

This paper is primarily to introduce the new ST device designed, built, and integrated indigenously, incidentally the first spherical tokamak in India to the broader fusion physics and engineering community with a special emphasis of device details, commissioning aspect and early plasma operations. The detailed physics design basis of the device and subsystems design details has been kept out of the scope of this paper and shall be reported subsequently.

## 2. Experimental System

IPR's journey towards a compact spherical tokamak fits largely into a well-established progression in global ST research, where major national programs first demonstrated the physics of spherical torus on small, low-cost devices before scaling up to larger machines with broader missions. At Culham, the START (Small Tight Aspect Ratio Tokamak) device was constructed as a modest, capacitor-bank-driven experiment that nevertheless provided the first clear demonstration of high-performance, low-aspect-ratio plasmas [3] and directly motivated the design and construction of the larger MAST facility. In the US, development of the NSTX program similarly drew on experience from smaller spherical torus experiments such as Pegasus and complementary ST devices like LTX [21-22], which helped to establish operating scenarios, engineering solutions, and control approaches suitable for the tight aspect ratio regime. In Japan, the QUEST program was preceded by the construction and operation of the CPD spherical tokamak [23-25], explicitly used as a testbed to develop and validate non-inductive start-up [26] and current drive without use of central solenoid and to prepare the operational basis for the larger QUEST device.

Historically, IPR's experimental tokamak program has focused on relatively high aspect ratio devices, both in copper and superconducting magnet technologies, so the move toward an ST represents a deliberate broadening of the program's parameter space spanning aspect ratios from less than 2 to greater than 5. The Aditya series copper-coil tokamaks [27,28] are conventional aspect ratio values around three, while the SST series superconducting machines [29,30] have operated at even higher aspect ratios above five, emphasizing long-pulse operation and steady-state technologies in the traditional tokamak configuration. In contrast, a spherical torus with aspect ratio $A < 2$ allows IPR to investigate the distinctive equilibrium and engineering features of very low aspect ratio plasmas—tight central column, strong vertical and shaping fields, fully non-inductive startup & current drive [31] and high-beta operation [32]—within a modest-cost, small-scale platform that is well aligned with how other programs have seeded their ST programs.

### 2.1 Device Parameters

The size of the new ST machine at IPR is set by several tightly coupled engineering constraints, reflecting both physics requirements and practical site limitations. First, the minimum bore radius of the central column must be sufficient to accommodate the central solenoid and the inner legs of the toroidal field coils, which is a defining geometric constraint in any spherical tokamak design. Second, the choice of toroidal-field coil system was guided by the decision to exploit an existing 2.45 GHz microwave source, which fixes the required on-axis magnetic field at approximately $B_0 \approx 0.876$ T to match the electron cyclotron resonance at the plasma center; this in turn sets a lower bound on the achievable major radius and coil dimensions for reasonable current densities and stress limits. Third, the overall linear dimensions of the device were constrained by the available laboratory space needed not only to house the tokamak itself but also to accommodate power supplies, vacuum systems, microwave and diagnostic equipment, and associated support infrastructure, forcing an integrated optimization of machine size, access, and layout.

The design considerations of each of the subsystems are detailed in the following sub-sections. Below is the table that summarizes the primary parameters of the new ST machine.

Table-1

| Parameter | Value |
|---|---|
| Major radius | 0.28 m |
| Minor radius | 0.16 m |
| Aspect ratio | 1.75 |
| Toroidal field (TF) coils | 6 coils × 3 turns |
| Central solenoid | 1 x 313 turns |
| Compensating coils | 3 pairs |
| Poloidal field (PF) coils | 3 pairs |
| Toroidal field ($R_0$) | ~0.1 T |
| Heating | Ohmic + ECR (2.45 GHz) |
| Phase-1 Operational Parameters | |
| Plasma current | ~30 kA |
| Plasma density | ~5 × $10^{18}$ m−3 |
| Electron temperature | ~50 eV |
| Ohmic plasma duration | 10 ms |

### 2.2 Vacuum vessel and support structure layout:

The vacuum vessel must accommodate the center stack, PF coils, diagnostic ports, vacuum pumping, and future maintenance. In the ST, the narrow central region makes even modest clearances important, so mechanical design has a direct influence on physics capability. The vessel should support accurate assembly alignment and magnetic reproducibility, since startup and control can be sensitive to geometric errors and stray fields.

Several vacuum vessel configurations were examined for ST during the design stage. Since spherical tokamaks inherently favor elongated plasma cross-sections, a conventional doughnut-shaped vessel was not considered suitable. For ease of fabrication concentric cylindrical geometry with top and bottom lids concept is adopted for the ST vacuum vessel, with the inner cylinder forming the central bore. The vessel wall thickness is selected as 6 mm with additional stiffeners optimizing the stress consideration of differential pressure, self-weight, and transient electromagnetic loading. To simplify fabrication, the top and bottom covers were designed as flat plates; however, this introduced sharp corners at the upper and lower junctions, necessitating an evaluation of peripheral stiffening ribs in both radial and vertical directions to improve structural rigidity and prevent buckling. Accordingly, the ST vacuum vessel is divided into an outer shell, an inner shell, and flat top and bottom plates. The detailed mechanical design considerations of the ST vacuum vessel are given elsewhere [33-Ranjith et al.]. A few major parameters are given in table-2.

Table 2 Major parameters of Vacuum Vessel

|  |  |
|---|---|
| Outer diameter (mm) | 1040 |
| Inner diameter (mm) | 170 |
| Wall thickness (mm) | 6 |
| Height (mm) | 772 |
| No of radial, top, bottom ports | 6 each |
| No of pumping ports (at bottom) | 2 |
| No of inner & outer limiters | 2 each |
| Volume of vacuum vessel (m3) | 0.62 |
| Resistance of VV (m-Ohm) | 0.05 |
| Inductance of VV (µH) | 0.048 |
| Time constant L/R (ms) | 0.9 |
| Mass of VV with port flanges and pumping lines (Kg) | ~700 |

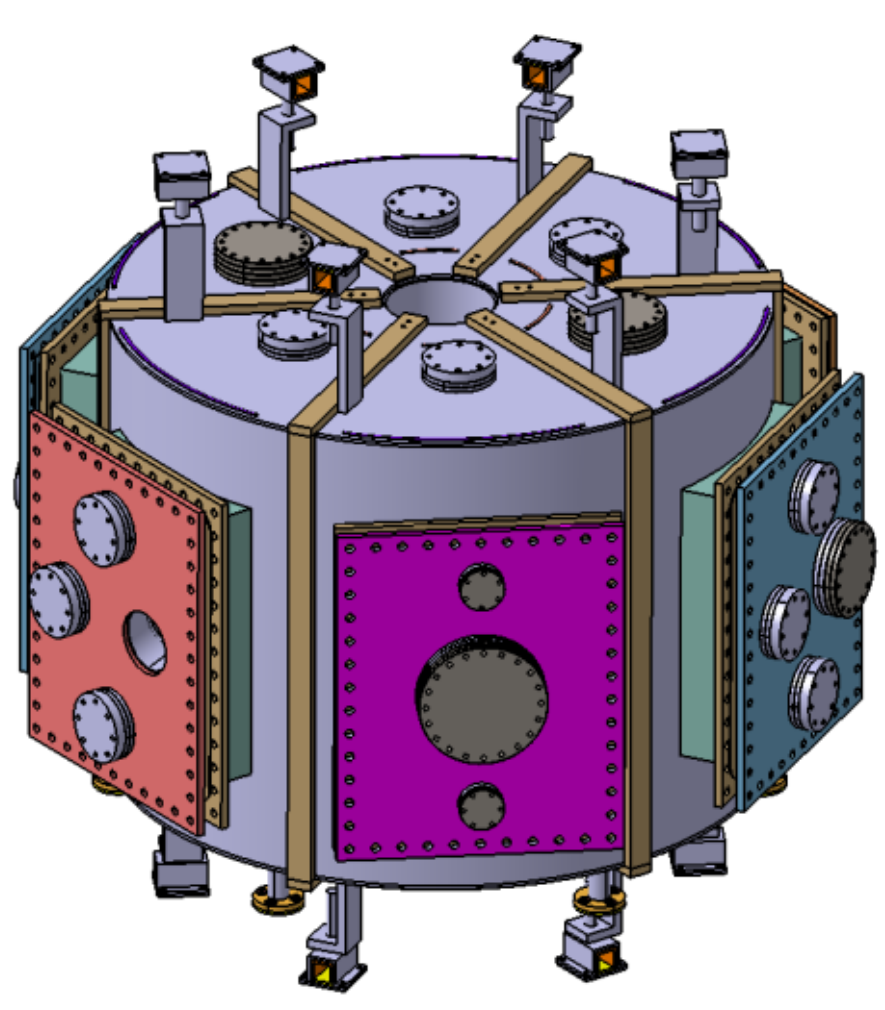

Figure-1: Vacuum Vessel structure showing radial ports, top ports and a few support structures for magnetic coils.

The vessel is evacuated by two set of turbo-molecular pumps with pumping speeds of 685 l/s backed by rotary pumps. Pneumatically operated gate valves isolate the vessel with the pumping system for maintenance, vacuum isolation, and wall conditioning purposes. Two vertical pipes with diameters of 153 mm are used to connect the TMP with vessel which is elevated 1.54 m above ground to accommodate the vacuum pumping system and to allow <6 mT magnetic field at the pump location during the plasma operation. Vessel pressure measurement was carried out by the universal gauge located on vessel and pumping line. An SRS make residual gas analyzer is installed on the pumping line to measure impurity gases partial pressure and quality of vessel vacuum right for plasma discharges. During the site acceptance tests, the vessel is baked for 150°C for 48 hours with the help of silicone baking tapes that resulted base pressure of $1x10^{-6}$ mbar. Subsequently, after a series of leak testing and careful sealing of the flange joints a leak rate of $<2x10^{-9}$ mbar.l/s is achieved and a base pressure of $1x10^{-7}$ mbar is obtained after 10 days of continuous pumping. The ultimate base pressure further reduced to $2.1x10^{-8}$ mbar after a series of wall conditioning. The baking system on vessel is presently not installed due to space availability between vessel and nearest magnetic coils installed. An integrated thin silicone based heating jacket is being considered that will enable the vessel baking that may further improve the ultimate base pressure.

### 2.3 Magnet system and center stack

The magnet system of the ST consists of toroidal field (TF), Central Solenoid (CS), three pairs of corrections coils (TR-1,2,3) and three pairs of poloidal field coils (PF-1,2,3) for startup, equilibrium, and shape control. The details of the coil position and dimensions are given in table-3. Because the central column is compact, structural, electrical, thermal, and insulation requirements are all tightly coupled. This makes the center stack one of the most critical design subsystems in any ST.

The TF coil system for the spherical tokamak comprises six coils, each with three turns, fabricated from copper sheets and assembled in two 90° legs. To reduce the toroidal field ripple to below 1% at the plasma centre, the outer legs of the TF coils are extended away from the plasma towards the low-field side (LFS). A key design constraint arises from the need to accommodate the inner legs within a 120 mm diameter inboard bore in order to maintain a machine aspect ratio of 1.75. Several TF coil design options were evaluated; these are described in detail by Ranjith et al.

One of the most challenging aspects of TF coil installation on ST machine is tightening the inner-leg TF coil joints within the limited space available at the centre bore. The complete TF coil assembly includes 40 interfaces, which contribute to the overall coil resistance. For 18 turns with a total conductor length of 84.4 m, using 30 × 10 mm² ETP copper, the calculated bulk coil resistance is approximately 4.92 mΩ, while the total resistance of the 40 joints is estimated at 0.34 mΩ. However, due to the restricted space at the inboard bore, applying the required torque to the joints proved difficult, leading to contact resistances an order of magnitude higher during initial trials. By employing specialized tools and refined joint-tightening procedures, the measured total joint resistance was reduced to 2.42 mΩ. Including the busbar resistance of 1.2 mΩ, the total resistance at the power-supply terminals is 8.04 mΩ, which lies within the allowable operating range of the power-supply voltage for delivering the current required to achieve $B_0$=0.0876 T.

The second most important magnetic coil required to produce required electric field for neutral breakdown and plasma current formation is ohmic transformer (OT) coil. This coil consists of a central solenoid (CS) wound in two layers. It has an inner radius of 0.06 m, a height of 0.942 m, and a total of 313 turns. The OT coil encloses the inner legs of the TF coils within the central bore of the vacuum vessel and is connected in series with three pairs of field correction coils (called TR coils) placed symmetrically above and below the mid-plane. The location of the field correction coils was optimized to minimize error fields inside the vacuum vessel due to the OT coil itself.

Table-3

| **Coil** | **R (m)** | **Z (m)** | **dR (m)** | **dZ (m)** | **$N_{turn}$** |
|---|---|---|---|---|---|
| CS | 0.07 | 0 | 0.020 | 0.942 | 313 |
| TR-1 | 0.13 | 0.466 | 0.018 | 0.018 | 7 |
| TR-2 | 0.564 | 0.55 | 0.01277 | 0.006385 | 2 |
| TR-3 | 0.67 | 0.5 | 0.006385 | 0.006385 | 1 |
| PF-1 | 0.18 | 0.466 | 0.01724 | 0.01724 | 4 |
| PF-2 | 0.47 | 0.47 | 0.00862 | 0.00862 | 1 |
| PF-3 | 0.642 | 0.33 | 0.022 | 0.022 | 4 |

Three pairs of poloidal field (PF) coils are installed on the device. At present, one pair is used to generate the vertical field required for J x B force balance and keep the plasma column at center. This pair has a diameter of 1.28 m and is located symmetrically at 0.33 m above and below the mid-plane. During operation, a maximum current of 4 kA is applied to this PF coil pair, producing a vertical magnetic field of 0.022 T at the minor axis of the toroidal vacuum vessel.

The vacuum vessel and magnetic field coils are supported by an outer structural frame consisting of six pairs (outer and inner) of stainless-steel pillars erected from the ground outside the vacuum vessel, together with additional central support. The pillars are interconnected by steel bars at multiple levels to withstand the mechanical loads and reduce vibration during operation. The central solenoid and the inner legs of the TF coils are mounted through this support structure from the center and below the vacuum vessel. Suitable insulating materials are used to electrically isolate the coils and the vacuum vessel from the support structure. The PF coil pair used in the present operation is located just outside the vacuum vessel and is supported by the outer structural frame. Details of the support structure design are described by Ranjith et al.

### 2.4 Power supplies system

Power-supply design strongly influences startup reliability, repeatability, and available operating space of a tokamak discharge. In small and medium tokamaks, current evolution is fast because the machine inductance is relatively low, which places tighter requirements on timing, triggering, and feedback during the earliest plasma phase. For an ST, where startup flexibility may already be constrained by central-solenoid limitations, reliable and reproducible power-supply sequencing becomes even more important. There are three power supplies for the first phase of ST operation, namely Ohmic Transformer Power Supply (OTPS), Toroidal Field Power Supply (TFPS), and Equilibrium Field Power Supply (EFPS for PF3 coil pairs) [34,35].

For plasma initiation and to achieve required loop voltage, Ohmic Transformer Power Supply (OTPS) is designed with the flexibility to achieve variable di/dt from 1~4 MA/s using double swing topology [34]. In this double swing operation, power supply operates in three phases while discharging to the OT coil viz., (a) OT coil charging phase, (b) fast di/dt phase, and (c) slower and longer di/dt phase. These three phases of OTPS operation are achieved using three capacitor banks and the switching devices. In first phase of operation, OT coil is charged to the peak positive current. The designed value of peak positive current is 1 kA to 5.5 kA. This positive peak current value is function of, (i) required time to achieve that peak value, (ii) pre-charging voltage of capacitor bank and, (iii) capacitance of the capacitor bank. By combination of capacitance and pre-charging voltage of capacitor bank the peak value and the duration of the peak current value can be varied. In the second phase, very fast rate of change of OT coil current (*di*/*dt*) is achieved using the selecting capacitance value of capacitor bank-2. The designed values of this fast *di/dt* are 1~4 MA/s. This phase produces the peak loop voltage to achieve gas breakdown and plasma initiation. In the third phase, slower *di/dt* and longer duration of pulse is required to sustain the loop voltage and plasma current. In third phase the current increases in reverse direction to the first phase of the current. The design values of peak current range from 1 kA to ~ 6.5 kA. In this phase, the peak value of current and pulse duration can be set by selecting the suitable values of pre-charging voltage and capacitance of the capacitor bank-3. All the capacitors of OTPS are Metalized Poly-Propylene (MPP) type and have very high fault current handling capabilities.

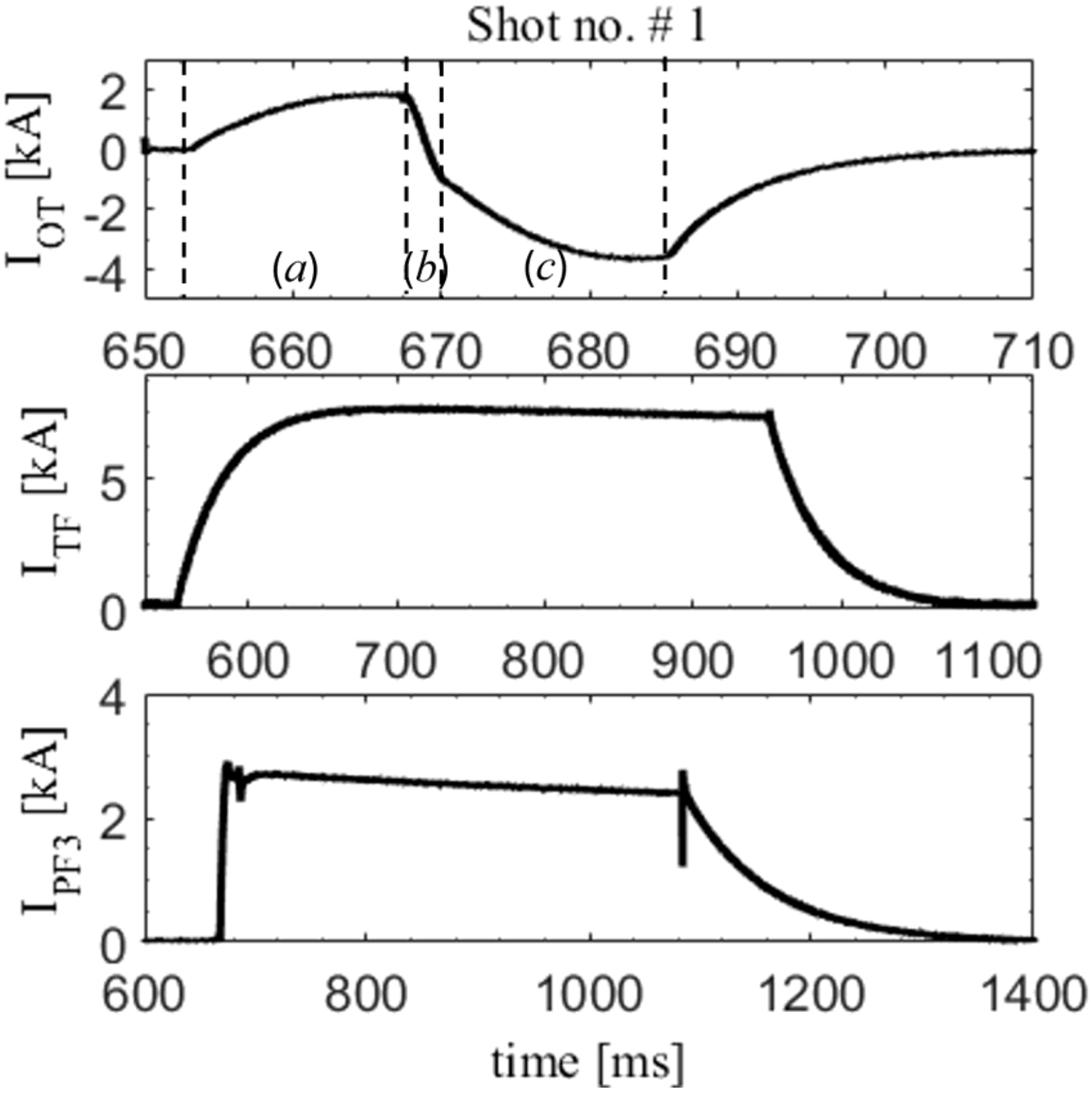

Figure-2: Time traces of coil currents in OT, TF, and PF3 during one of the typical discharges at a reduced parameter of the power supplies. Various phases of OT coil currents (a) charging, (b) fast discharging, and (c) slow discharging phases are indicated by broken lines.

Toroidal Field Power Supply (TFPS) provides strong toroidal fields to confine the plasma inside the vacuum vessel. The TFPS is Super-Capacitor based compact power supply [35]. Super-capacitor bank provides very high-power density and lower size footprint. The super-capacitor bank used in TFPS is rated as 608 F and 153 V. The design value of TFPS peak current is ~13 kA, which is estimated to achieve 0.15 T magnetic field at the major radius of the tokamak. The design value of the flat-top TF coil current is 100 ms to 250 ms. The allowable droop in peak TF coil current is designed at approximately 1% for 50 ms time duration. The required TF coil current ranges from 1 kA to 13 kA, is settable by pre-charging voltage of the super-capacitor bank. Thyristor stacks have been used as switching devices and, to achieve a variable pulse duration, thyristor commutation circuit is used.

Equilibrium Field Power Supply (EFPS) powering PF-3 coil is a two-stage capacitor-bank based power supply. The current rating of the EFPS ranges from 1 kA to 4.5 kA. There are two types of capacitor banks used in the EFPS viz. (a) MPP capacitor bank and, (b) Super-capacitor bank. EFPS works in two phases of operation. In first phase, high voltage capacitor bank provides fast *di/dt* ranging from 1 MA/s to 3 MA/s. This *di/dt* range depends on the selection of capacitance value of MMP capacitor bank. The provision for the selection of tap in this capacitor bank provides flexibility in *di/dt* selection. The super-capacitor bank drives the second phase and provides longer pulse durations. The pre-charging voltage provides flexible peak current values in both phases. To control the coil current discharge, thyristor is used as switching device and, provides the variable pulse duration through its commutation circuit. A typical waveform of all three power supplies operating on coils is shown in figure 2.

**2.5 Gas feed and wall conditioning system:**

The first plasma is often limited not by ideal equilibrium considerations but by wall condition, impurity influx, recycling, and visible-light radiation. For this reason, vessel cleaning and glow discharge cleaning becomes very important. During the vacuum vessel installation, the baking is carried out by isolated heating tapes, however, after all coil installations, the required space for baking tape becomes more restrictive, at places less than 7mm. Hence a thin wall special heating jacket is being designed and considered for the future. To prepare the machine for the first plasma, no suitable baking system was available. Therefore, the two vertical rod-based Glow Discharge Conditioning (GDC) systems have been designed and installed diametrically opposite toroidally on the top plate of VV and their position is pneumatically controlled remotely. A variable DC power supply up to 1000 V/2A powers the GDC rods made of SS304L and works as anode, while the vacuum vessel works as cathode to accelerate ions to the wall that aids in wall cleaning. GDC method is one of the primary tools used in tokamaks for removing impurity sources containing oxide, carbide and other impurities. The resulting low temperature plasma during GDC generates ions and metastable species that bombard the vessel wall surfaces, desorbing and pumping away trapped gases such as hydrogen, water vapour ($H_2O$), oxygen ($O_2$), carbon monoxide (CO), carbon dioxide ($CO_2$), and hydrocarbons (e.g., $CH_4$). These released impurities are detected by the RGA described in previous section. In the present experiment, GDC is operated 0.1–0.4 A/m$^2$ current density and a pressure of 1x10$^{-2}$ to 1x10$^{-3}$ mbar. More details of the GDC design, installation and test results are described by Khan et. al. A hydrogen gas feed system with piezoelectric valve is installed to inject Hydrogen gas into the tokamak to produce Hydrogen plasma. A hot palladium membrane-based Hydrogen purifier system is installed to inject up to 7N pure Hydrogen gas into the vacuum vessel during plasma discharge. Separate provision is made to install various other gases like Argon, Neon, Helium for various wall conditioning experiments (figure-3).

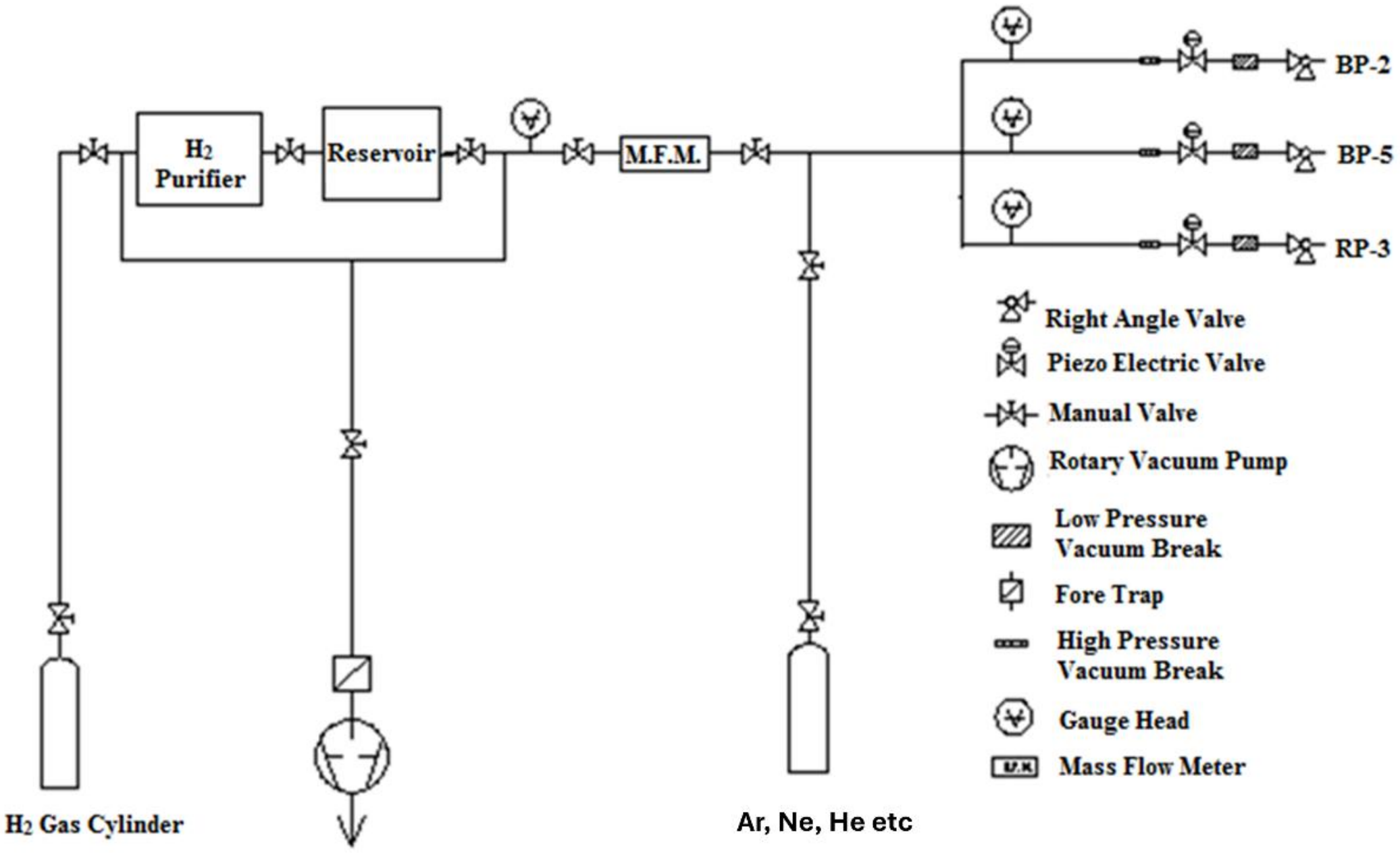

Figure-3: Line diagram of gas feed system for ST machine used for prefilling working gas during plasma experiment or wall conditioning.

### 2.6 Machine Diagnostics:

As a commissioning-grade ST device, a core set of diagnostics sufficient to establish whether the machine has formed and sustained a tokamak plasma are integrated. The diagnostics include internal Rogowski coils, one turn flux loops, $B_\theta$ magnetic probes [Sameer et. al], Visible and NIR spectroscopy including impurity measurement [Neelam et. al.], vessel current monitors by Fiber Optic Current Sensor (FOCS) [Santosh, Sheetal et. al.], Visible camera, movable Langmuir probe, one channel Bolometer, X-ray monitoring system [Shishir et. al.], vessel temperature monitoring, Rogowsky coils for all machine coils [Navi, Sameer et al.],

Residual Gas Analyzers and fill pressure gauges. Arrangement of all these diagnostics on the machine is shown in figure-4, while the position of flux loops that measures the loop voltages are given in table-4.

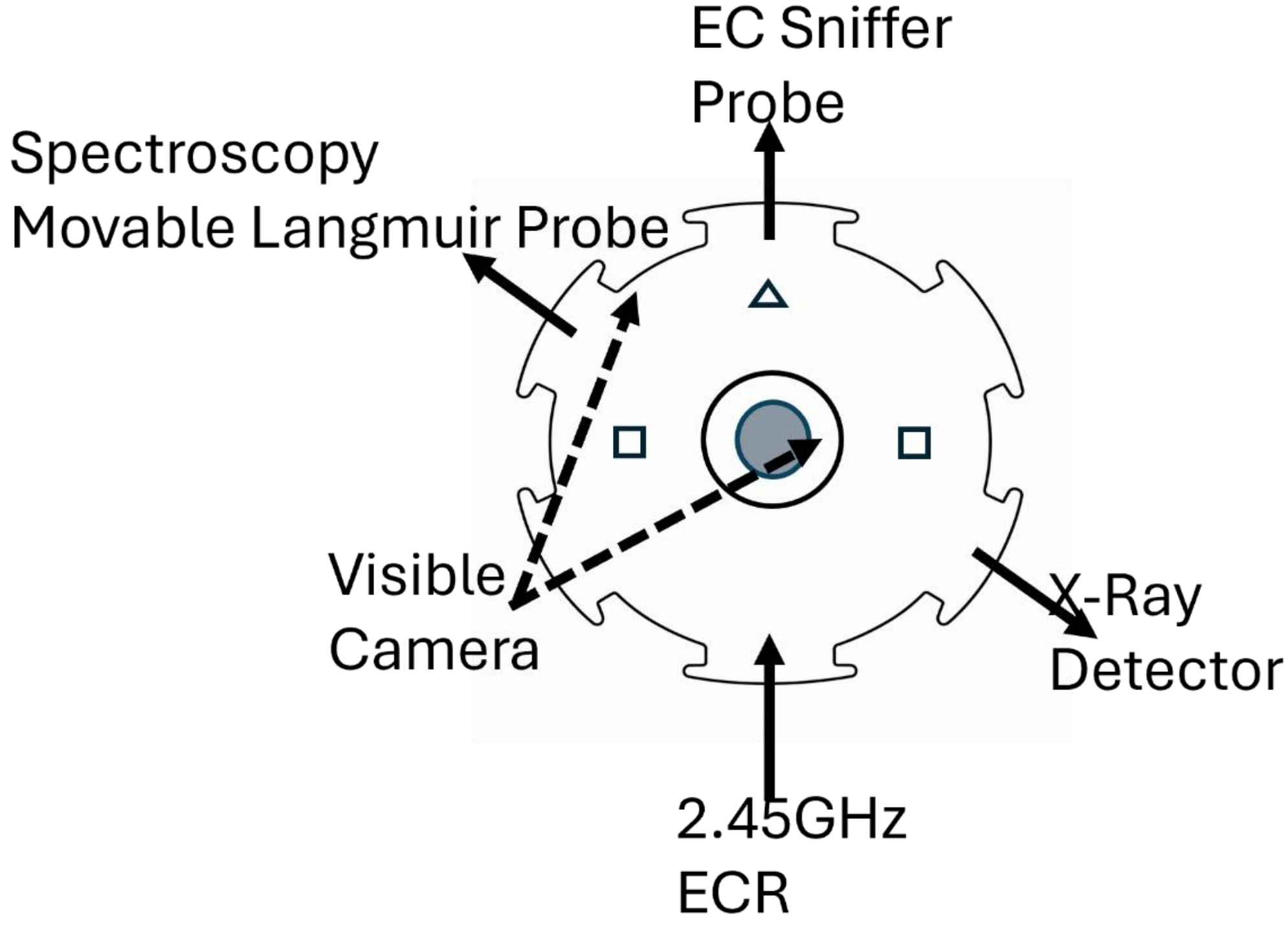

Figure-4: A midplane toroidal view of the ST vacuum vessel showing location and field of view of various diagnostics installed during first plasma commissioning. Vessel radial port numbering starts at 1 from top of the figure and counterclockwise. The two square boxes on machine top port indicate location of GDC rods and the triangle indicates one channel bolometer diagnostic.

Table-4

| **Flux Loop No.** | **R (m)** | **Z (m)** |
|---|---|---|
| 1 | 0.092 | 0.380 |
| 2 | 0.092 | -0.380 |
| 3 | 0.505 | -0.250 |
| 4 | 0.505 | 0.250 |
| 7 | 0.511 | -0.389 |
| 6 | 0.173 | -0.389 |
| 5 | 0.173 | 0.389 |

| 8 | 0.511 | 0.389 |
|---|---|---|

A visible survey spectrometer system has been installed on the ST tokamak to record line-integrated plasma emission. Plasma light is collected through a collimating beam probe mounted on the viewport of radial port 2. The probe defines a viewing chord across the horizontal mid-plane and is coupled to a 10 m long optical fiber with a 1 mm core diameter and numerical aperture of 0.22. The fiber is connected to a three-channel visible spectrometer, with each channel covering 350–900 nm and equipped with a 3048-pixel linear CCD detector. The spectral resolution varies from 2.3 to 2.6 nm over the operating wavelength range. Data acquisition is initiated using an external trigger synchronized with the tokamak loop voltage.

Spectra were recorded during glow discharge cleaning (GDC) and plasma discharges. Several atomic emission lines and molecular bands were identified, including the prominent $H_\alpha$, $H_\beta$, and Fulcher bands in the wavelength range 601–635 nm, as shown in Fig. 5.

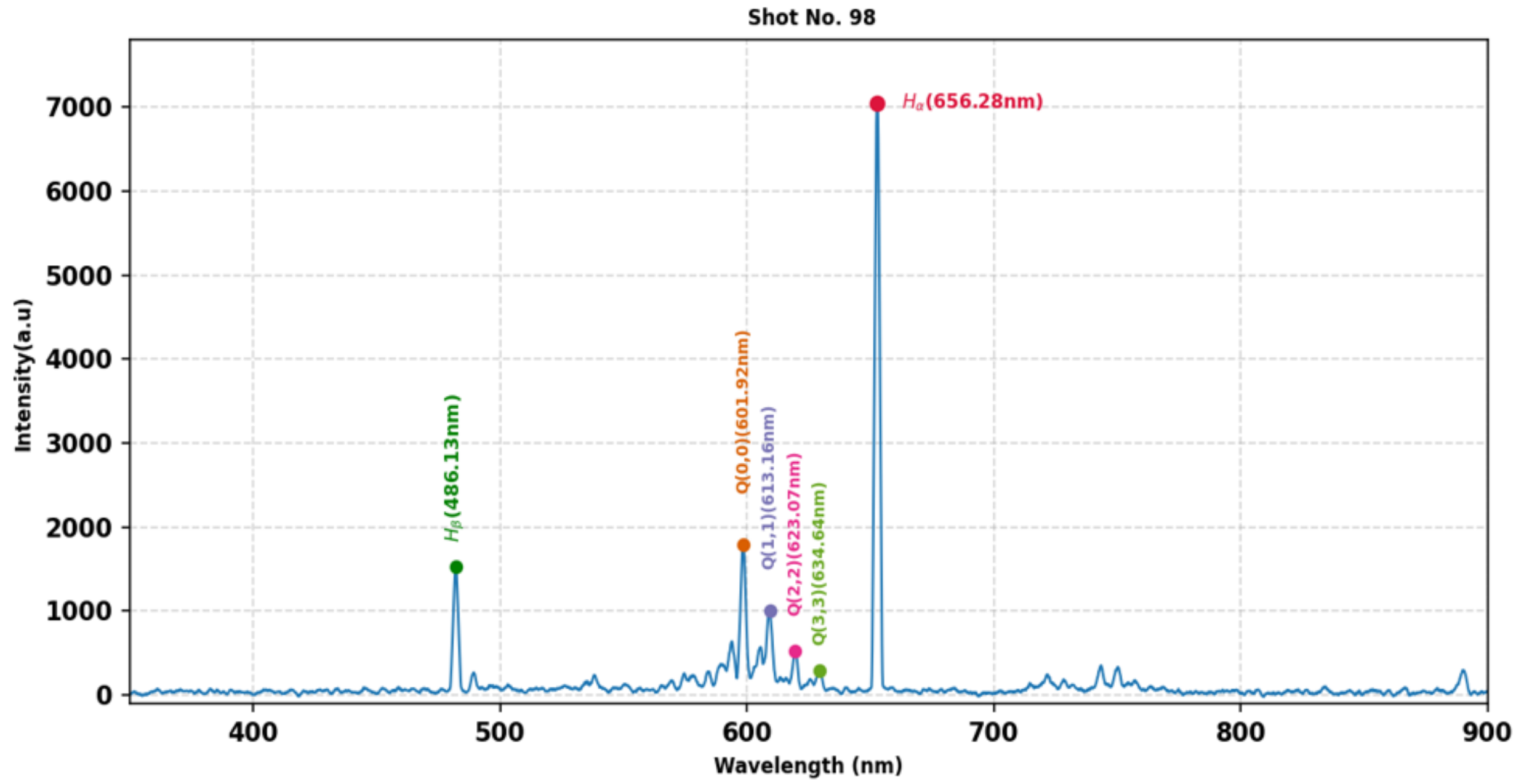

Figure-5: A typical spectral lines during one of the open field line plasma discharge recorded by the survey spectrometer.

Additionally three photomultiplier-tube (PMT) diagnostic systems have been installed on the machine to monitor the temporal evolution of the $H_\alpha$, $O^{+1}$ , and $C^{+2}$ emission lines. Each system consists of plasma-side and detector-side collimating lenses, an optical fiber, an interference filter, a PMT detector, a biasing circuit, a signal-conditioning module, and a data-acquisition channel. The plasma-side lens is mounted on a 63 CF optical viewport at radial port 2 and defines the viewing field inside the plasma. The collected radiation is transmitted through a UV–visible optical fiber having a 1 mm core diameter and numerical aperture of 0.22. Both collimating lenses have a focal length of 19 mm and a diameter of 11 mm. Narrow-band interference filters with 1 nm bandwidths are used to select $H_\alpha$ at 656.3 nm, $O^{+1}$ at 441.5 nm, and $C^{+2}$ at 464.7 nm. The detector consists of a 28 mm head-on PMT, a high-voltage power supply, and a low-noise amplifier. Its photosensitive area is 25 mm, with an operating spectral range of 300–650 nm and a bandwidth from DC to 20 kHz. The current-to-voltage conversion factor is 1 V/$\mu$A. The detector gain is controlled by a bias voltage in the range 0.4–1.2 V. The PMT output is passed through a signal-conditioning module, which provides electrical isolation, reduces noise associated with ground loops, and supplies the conditioned signal to the data-acquisition system.

Figure 6 shows typical ST discharge parameters, including ECR forward power, ohmic current, and TF coil current, together with the time evolution of the $H_\alpha$, $O^{+1}$ , and $C^{+2}$ signals. Weak signals are observed when the ECR power reaches its maximum and when the TF current begins to rise. The emission increases significantly during the rapid variation of the OT current, when the loop voltage is established. The signals subsequently decrease and approach zero as the TF current begins to fall.

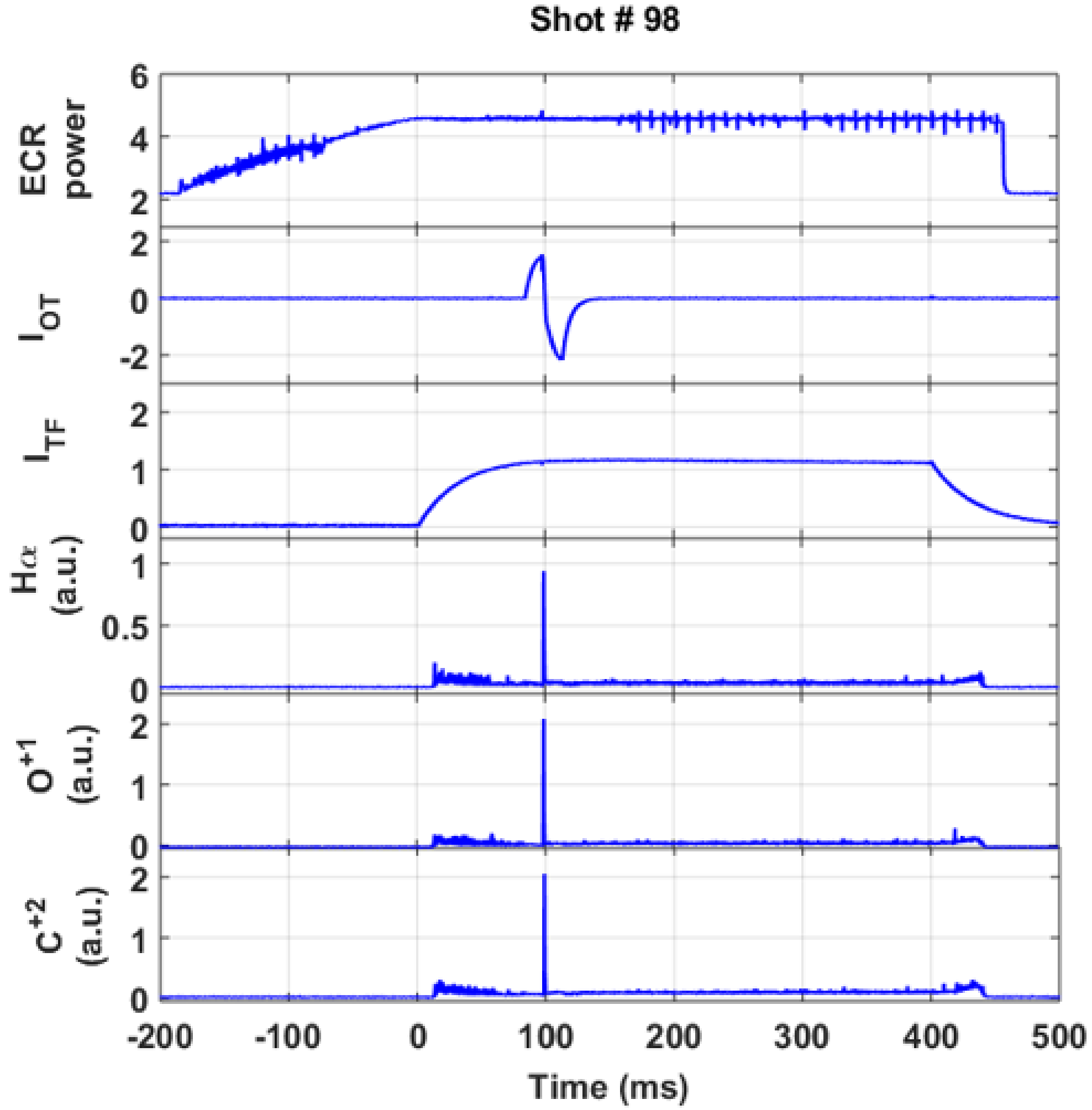

Figure-6: Various emission lines detected during the ECR phase and Ohmic phase of a plasma discharge. [This figure has to be merged with the ECR FP,RP from Jagabandhu and other diagnostics signals.]

**2.7 Electronics and Signal Conditioning System:**

Signal conditioning electronics has been designed and developed to meet the requirements of the diagnostic systems installed in the machine. Depending on the diagnostic principle, these sensors provide either voltage or current output signals with widely varying amplitudes and signal characteristics. The front-end electronics conditions these signals and scales them to the appropriate input range of the data acquisition (DAQ) system while maintaining adequate signal conditioning and dynamic range.

For example, the bolometer employs an AXUV photodiode as the sensing element, which produces a current proportional to the incident radiation. An I–V converter with a typical transimpedance gain of 1 V/mA has been designed to convert the photodiode current into a voltage signal suitable for acquisition by the DAQ system. Similarly, the loop-voltage signal typically has an amplitude of approximately 30 V. A 10:1 resistive attenuator is therefore employed to reduce the signal to a level compatible with the DAQ input range.

A further important consideration in the design of the magnetic diagnostic interfaces is their susceptibility to high common-mode voltages. Magnetic probes, including Rogowski coils and Mirnov probes, can experience significant common-mode voltage pickup due to their proximity to high-voltage and high-current components of the machine. To mitigate this effect and ensure reliable signal transmission and measurement, high-common-mode-voltage isolation amplifiers are employed at the first stage of the signal-conditioning chain for the magnetic diagnostics. An optocoupler-based isolation stage provides galvanic isolation between the electronics and the centralised data acquisition and control system (described in following sub-section), thereby improving common-mode rejection and maintaining zero ground loops.

### 2.8 Data Acquisition and Control System:

The ST Control System (STCS) is the supervisory framework of the ST machine that coordinates the operation of interconnected plant systems such as vacuum, magnets, power supplies, gas injection, and ECR system. It provides centralized control, monitoring, communication, and synchronization to ensure reliable execution of experimental sequences, including plasma initiation, discharge, and shutdown. The STCS follows a modular, two-layer architecture comprising a supervisory control layer for shot sequencing, timing, operator interaction, and overall coordination, and a plant system control layer consisting of dedicated local controllers for real-time control, data acquisition, monitoring, and interlock functions. The system also supports fast and slow diagnostic data acquisition, data processing, archiving, and post-discharge analysis while continuously monitoring machine status and responding to abnormal or emergency conditions. This layered architecture provides a flexible and scalable platform for integrating existing and future ST plant systems and diagnostics. The architecture diagram shown in Figure 7, presents main functional components of the STCS for ST device.

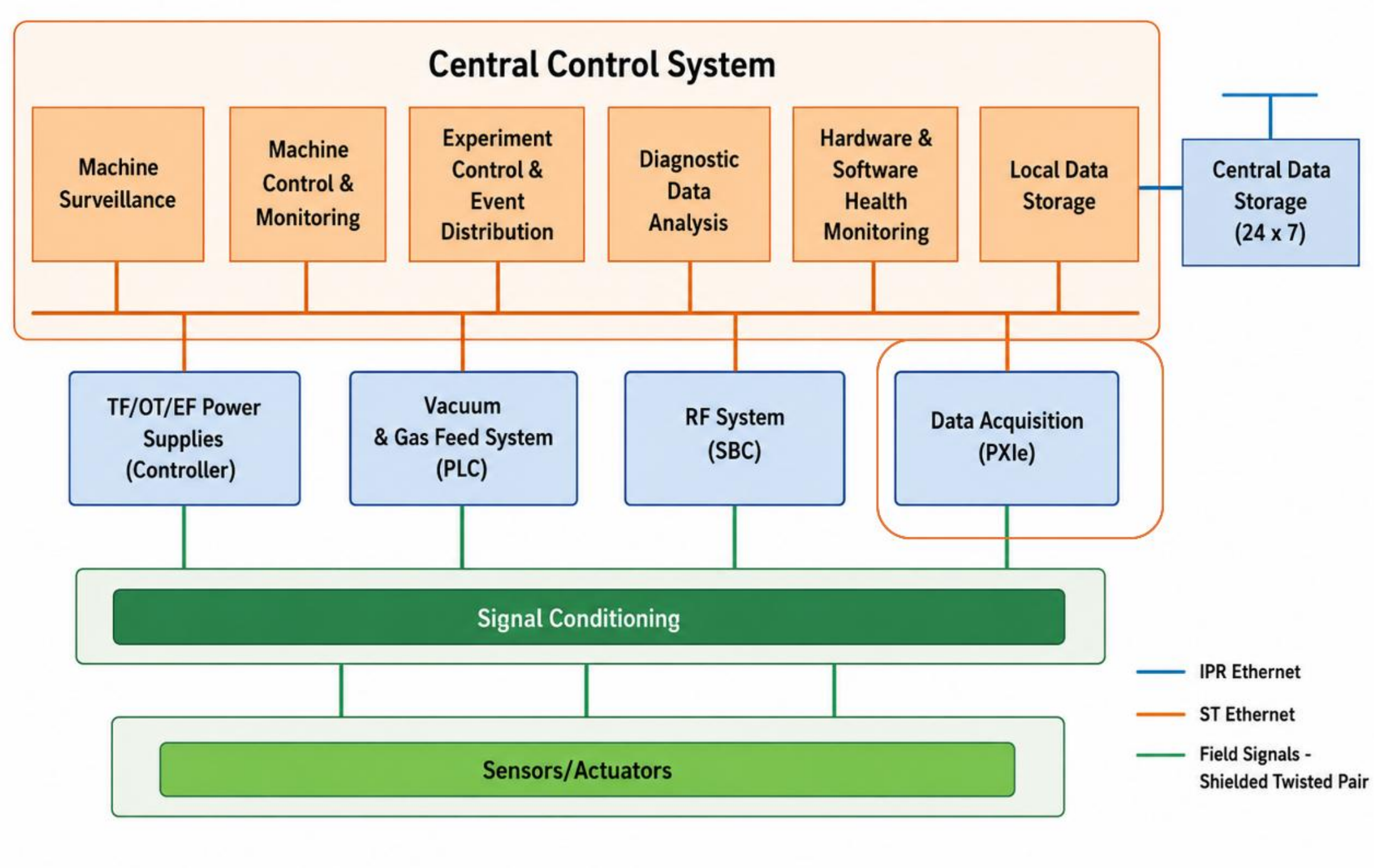

Figure 7: ST Control System Architecture

The ST Data Acquisition System (ST-DAS) is a modular PXIe-based platform developed for acquiring and recording experimental data from plasma diagnostics and other plant subsystems. It receives signals from front-end electronics through dedicated signal interfaces, digitizes those using appropriate PXIe modules, and transfers the acquired data to local as well as remote storage systems. The ST-DAS software configures and controls the acquisition hardware according to the experimental requirements, supporting different acquisition modes, including event-triggered operation. The system comprises PXIe chassis, controllers, data acquisition modules, and associated signal-conditioning and interface hardware, providing a flexible and scalable solution for synchronized acquisition of diagnostic and machine-related signals. The initial number of signals and their corresponding sampling rates configured in the ST-DAS are shown in Table-5.

Table-5 (ST- Data Acquisition System Sampling Rates)

| Sampling Rate (Hz) | No. of Signals |
|---|---|
| 10 kHz or lower | 5 |
| 50 kHz | 23 |
| 100 kHz | 27 |
| 200 kHz | 12 |
| 1 MHz | 2 |
| Total | 69 |

**2.9 ECR System:**

A 2.45 GHz/6kW CW magnetron based RF source is integrated to the ST machine for the purpose of preionization and EC heating experiment. The magnetron head along with a water cooled isolator is housed in a rack mounted module 3 m away from the ST machine followed by WR-340 rectangular waveguide, directional coupler for forward and reflected power measurement, a DC break to isolate RF ground, a vacuum break, and a waveguide-based launcher ($TE_{11}$ mode) to inject RF power to the vessel. A local control unit supervised by the STCS controls the On-OFF timing and RF power of the magnetron. The complete system is tested for vacuum and RF integrity before the machine commissioning. A block diagram of the ECR system along with the elevation diagram is shown in figure-8.

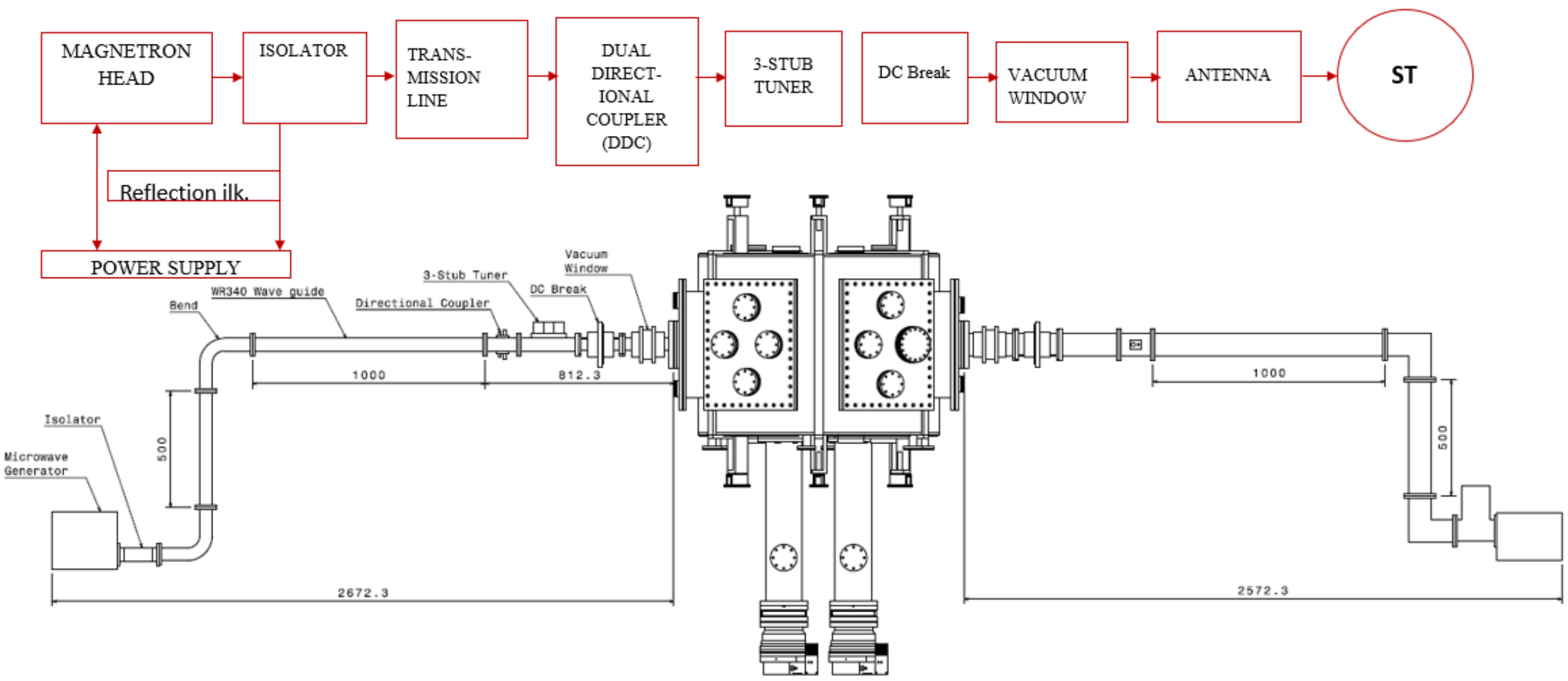

Figure-8: The line diagram and elevation view of the ECR system commissioned on the ST machine showing various parts of the system.

**3. Machine Commissioning and First Plasma:**

Commissioning a tokamak comprising multiple interdependent subsystems constitutes the critical first step toward achieving first plasma operation. In the case of the ST, this process commenced with the individual validation of subsystems, including vacuum integrity, glow discharge cleaning (GDC), power supplies (initially on dummy loads and subsequently on the ST magnetic coils), electronics modules with dummy signals, diagnostics, and the electron cyclotron resonance (ECR) system. Each subsystem was first tested under local control before being integrated into the central ST Control System (STCS), with all data archived in the ST Data Acquisition System (ST-DAS). During the individual magnetic coil charging experiments, the magnetic fields produced by the TF, OT, and PF3 coils were measured and verified. Following the successful commissioning of individual subsystems, an integrated commissioning phase was initiated, wherein subsystems were sequentially brought online in conjunction with the STCS. During this phase, noise floors for all diagnostics were verified, and data were acquired with all magnetic coils and ECR systems in operation. The following sections detail the outcomes of these commissioning activities.

**3.1 Magnetic Field Measurements:**

Toroidal and vertical magnetic field measurements were performed using a Gauss probe mounted on a translation stage for the selected coil configuration. The toroidal field was measured just outside the radial port under present operating conditions, with the TF coil current pulse applied for approximately 0.4 s and reaching a peak value of 8.3 kA corresponding to $B_0$ = 876 G required for fundamental EC resonance at plasma center. The radial profile of the toroidal field, however, was obtained at a reduced flat-top current of 0.5 kA by translating the probe on a shot-to-shot basis for ease of experiment. The resulting $B_\phi(r)$profile exhibits a reasonably good $1/R$ dependence, as expected. The vertical magnetic field ($B_z$) was measured with the probe head aligned in the vertical direction, and a time trace was recorded using 0.1 kA current in the PF coil pair supplied by a DC power source. In the present device, only one of the three PF coil pairs i.e. PF-3 is utilized during initial operation. The PF current ramps to flat-top in about 0.5 s, during which the changing magnetic flux induces a toroidal current in the continuous vacuum vessel; a fraction of this induced current was monitored via a copper bar mounted on the radial port using a current sensor. The induced current reverses sign with the change in slope of the PF current during the transient phase and decays to zero at flat-top, at which point the measured vertical field arises solely from the PF coil current. During the current ramp-up, the changing flux is expected to induce a toroidal current in the vacuum vessel, estimated indirectly from the magnetic field measurements to be approximately 15 kA at the outboard side. Details of the measurement procedure and this estimation are provided in the reference [36-Shekar at al.]

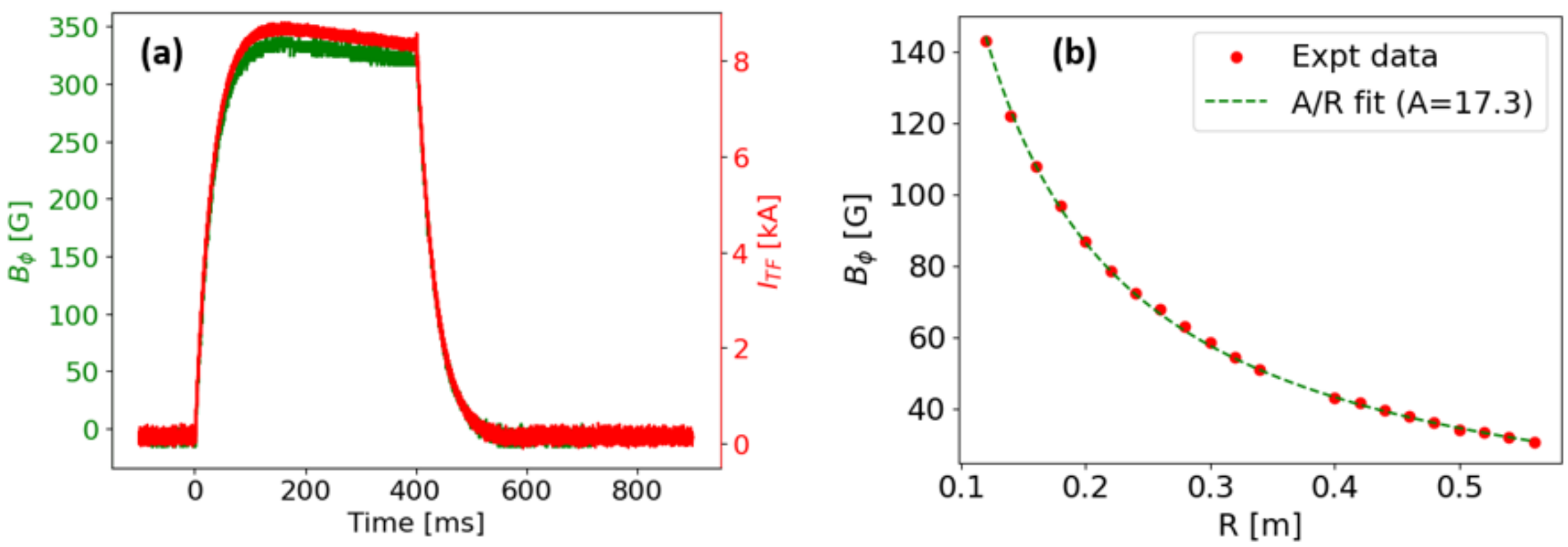

Figure-9: (a) Time trace of TF coil current and measured $B_t$ through a Gauss probe outside the vacuum vessel. (b) Radial variation of $B_t$ ($R$) obtained with reduced $I_{TF}$=0.5kA. [To be rebuilt in greyscale]

Similarly, the OT coil and PF-3 coil are also separately tested first with a low current DC power supply and later with the OTPS and EFPS respectively.

**3.2 Integrated Commissioning:**

Magnetic Field Coil System: Toroidal Field (TF), Ohmic Transformer (OT), and Poloidal Field (PF-1,2,3) coils are connected to their busbars. The OT and TF inner and outer leg joints were integrated under tight space constraints with minimum contact resistance to achieve desired value [give a figure of joint resistance variations]. Coil-to-vessel and coil-to-structure insulation resistance exceeds 1 GΩ, suitable for high-voltage operation. Coil Power Supply System: Capacitor-bank supplies (TFPS, OTPS, EFPS/PF-3) are integrated with the respective coils via HVDC bus bars. Each supply was individually tested, then operated together under central triggering system. Each power supplies are tested on the machine on coil separately and sequentially. Later combined operation to ensure the integrated timing sequence and operation. All diagnostic signals are conditioned for electrical isolation, gains etc. and then acquired with a PXIe based system. Cold tests, noise immunity checks (with coils charged), and integrated operation with the STCS confirmed performance and reliability of every sub-system associated with the tokamak.

Initially, the GDC plasma was struck at a higher pressure of > $1\text{x}10^{-2}$ mbar which is subsequently reduced to $5\text{x}10^{-4}$ mbar over a minute, where GDC sustains successfully for several hours. Later on, to ensure minimal gas loading, the GDC was initiated at lower pressure (~$5\text{x}10^{-4}$mbar) using the 2.45 GHz ECR source and sustained for longer period. Impurity species like $H_2O$, $O_2$, $H_e$, $C$ etc are reduced by an order of one to two and this brought the ultimate vacuum of the tokamak to $2.1\text{x}10^{-8}$ mbar from earlier $1\text{x}10^{-7}$ mbar in the installation phase.

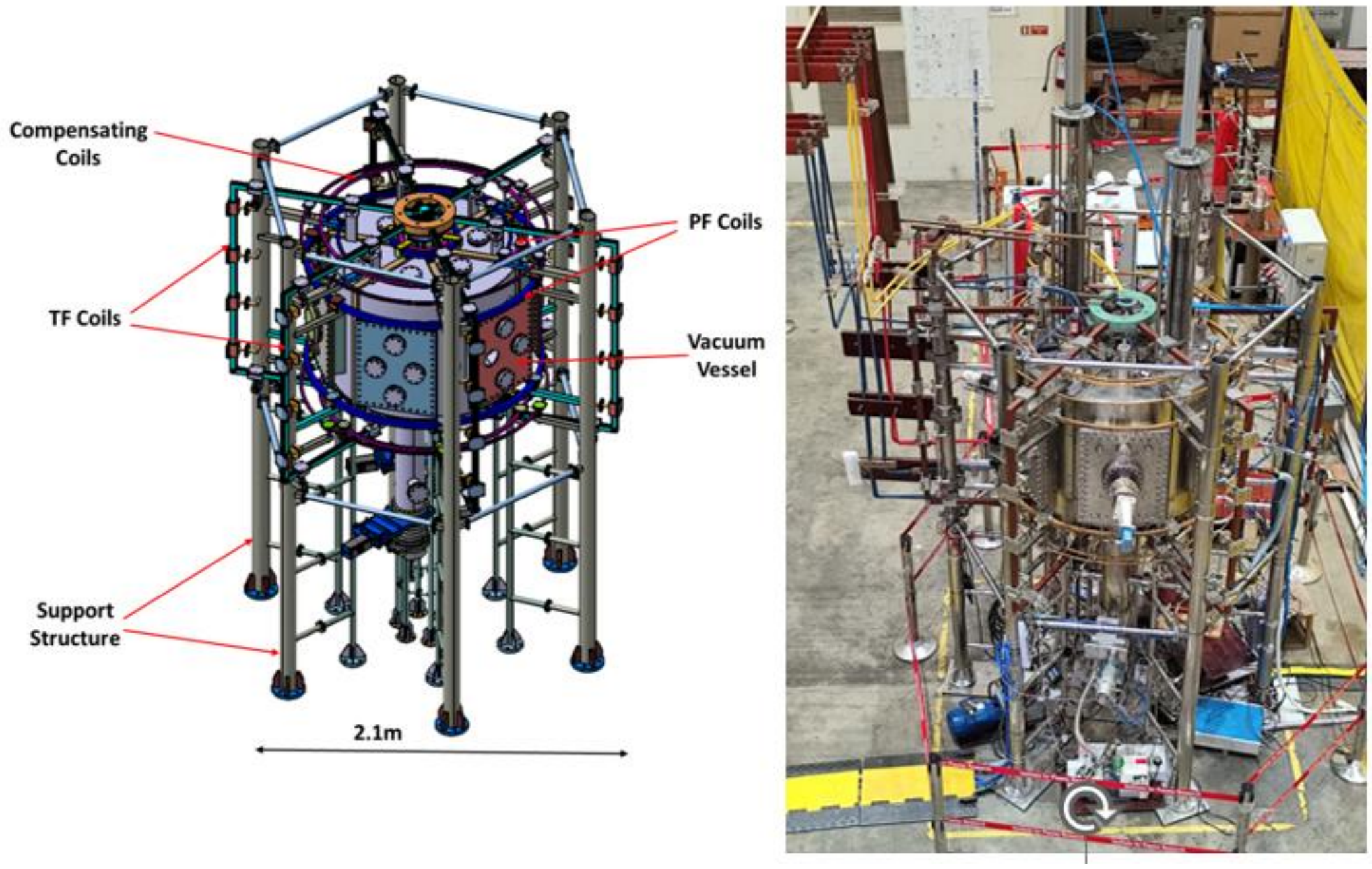

Figure-10: A CAD model with a scale and commissioned machine is shown side by side for comparison.

### 3.3 First Plasma Experiments:

After series of engineering and system level functioning validation, the ST machine is commissioned achieving its first plasma on 11 December 2025. The STCS operated in pre-programmed mode, injecting hydrogen gas into the vacuum vessel followed by sequential application of magnetic field coil currents as shown in figure-11.

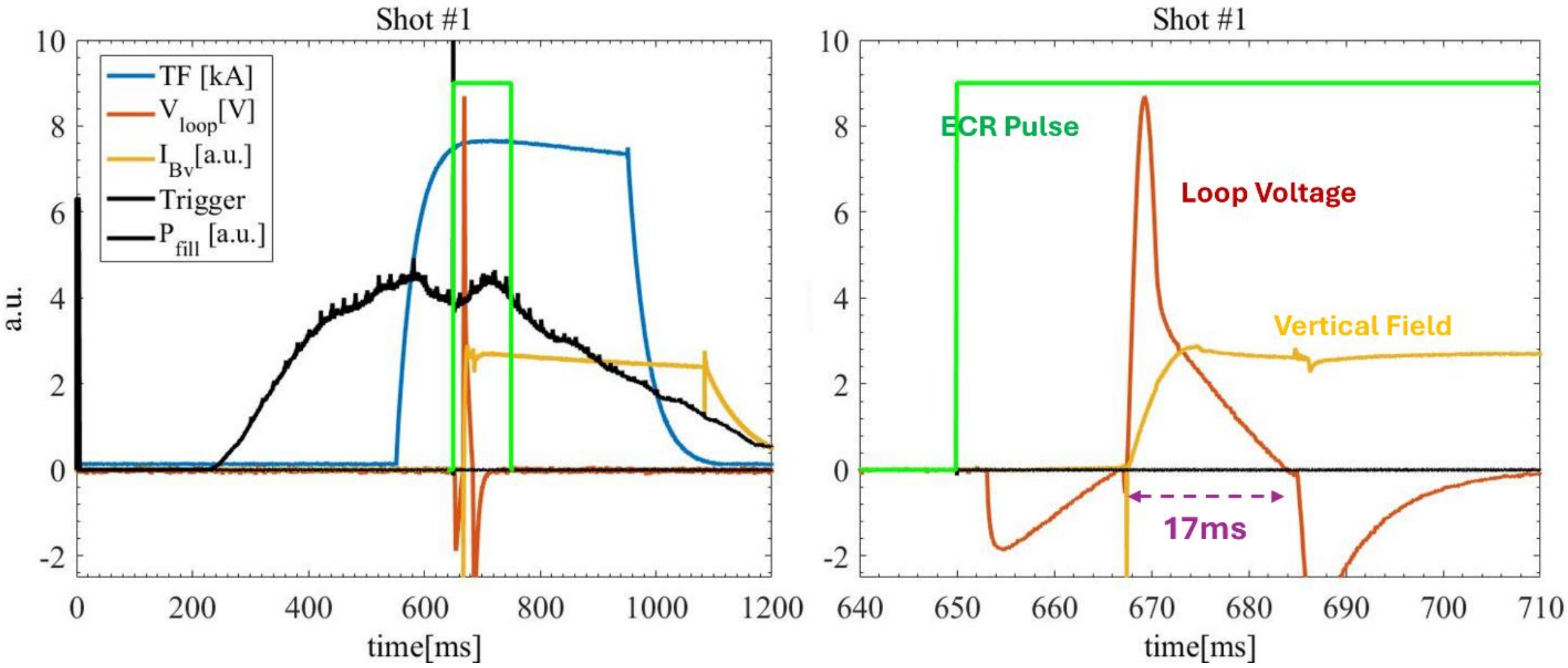

Figure-11: Time traces of important operating parameters for the first plasma commissioning of ST device. The left side figure shows the timing of master trigger followed by gas pressure, TF current, ECR pulse, OT

& PF3 (I_BV) currents as time progresses in this sequence. The right-side figure is an enlarged section of the left side figure during the Ohmic breakdown phase.

The sequence of operation is as follows: The STCS fires the master trigger once all the capacitor banks of three power supplies are fully charged to the set values. The master trigger starts with the acquisition of data and sequentially generates and send triggers to the individual subsystems as per the predefined sequence and time delays. The piezo-electric valve gets the first trigger, and gas pressure starts rising from 200 ms. Once the peak pressure is achieved, the TF field is applied, and it takes approximately 50ms to reach the required flattop value. The ECR pulse is applied, in this case 600 W that produces the preionized plasma. Loop voltage is applied to this plasma with vertical field to keep the plasma column in equilibrium. Plasma was successfully produced on the very first attempt. High-speed visible imaging captured the plasma formation, showing filament structures aligned with the toroidal magnetic field indicating open field line configuration. All diagnostic systems performed as expected, confirming hydrogen neutral breakdown forming a ring-shaped toroidal plasma column with excited impurity lines like $H_\alpha$, $C^{2+}$, $O^{1+}$ in the visible range.

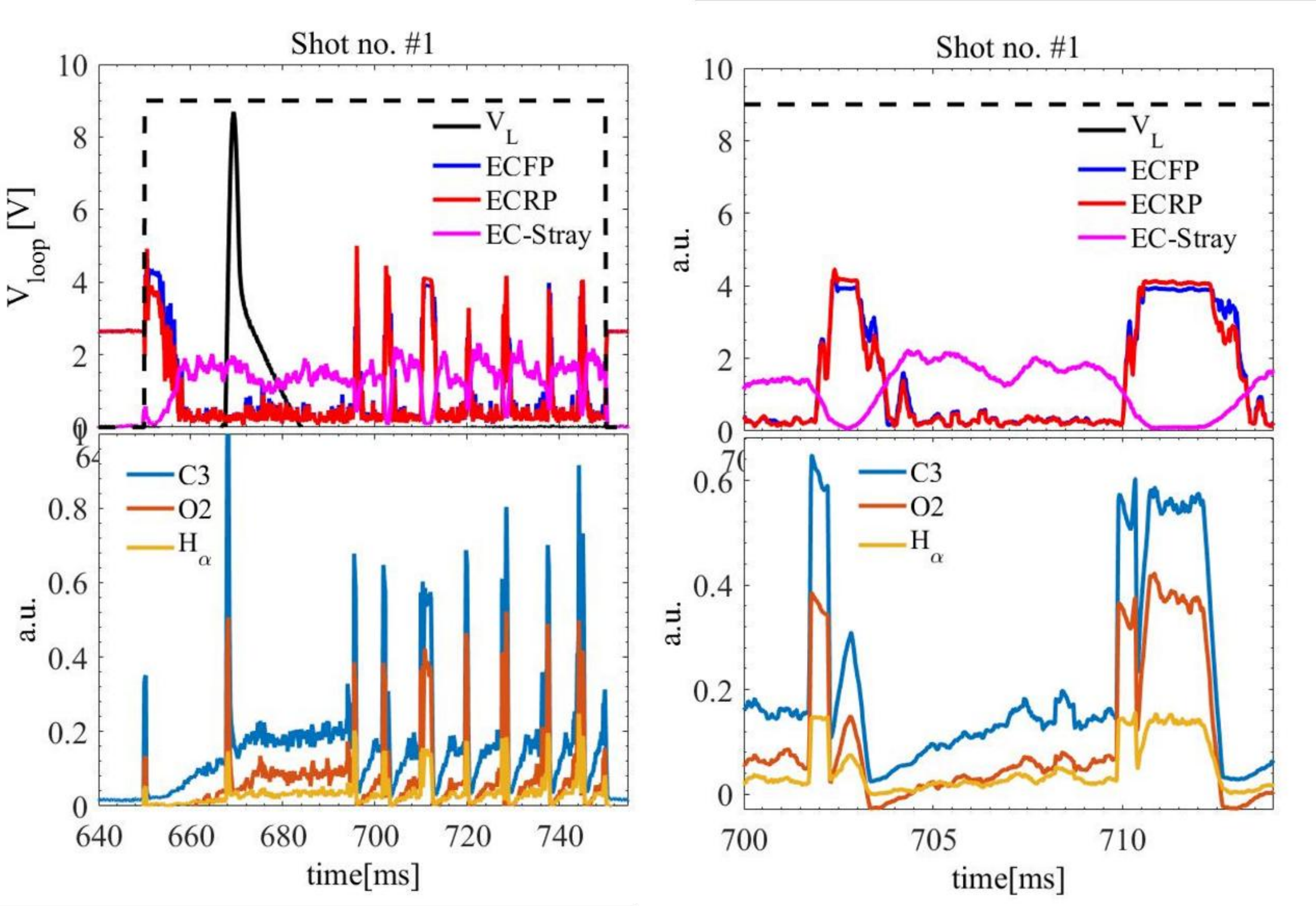

Figure-12: Time traces of loop voltage, EC forward and reflected powers, EC unabsorbed power and impurity radiation signals during the Ohmic and EC phase of the first plasma shot shown in figure 11. The right side figure is enlarged to show the periodic absorption of EC waves.

In Shot #1 the temporal evolution of the EC signals and spectroscopic lines reveals a clear correlation between absorption quality and plasma–wall interaction under open-field-line conditions. When the EC stray power decreases substantially (figure 12, interval ~708–712 ms), the EC forward power remains high and the $O^{+1}$, $C^{+2}$ and Hα intensities rise markedly, indicating strong plasma–wall interaction. At approximately 712 ms the EC absorption begins to weaken: the forward and reflected power drop while the stray radiation starts to increase, concurrently the impurity and hydrogen line intensities fall rapidly. This could be due to a reduction in local density and/or electron temperature that lowers the EC absorption efficiency, thereby increasing the residual stray radiation and diminishing the intensity of plasma–wall interactions. Because no plasma current is formed in this discharge and the magnetic configuration remains open-field-line, absorption is expected to be multipass and sensitive to wall conditions, which are still relatively poor in these early discharges. A more detailed investigation of the density and temperature evolution, together with ray-tracing of the EC beam, is required to quantify the absorption fraction and the associated wall-loading mechanisms.

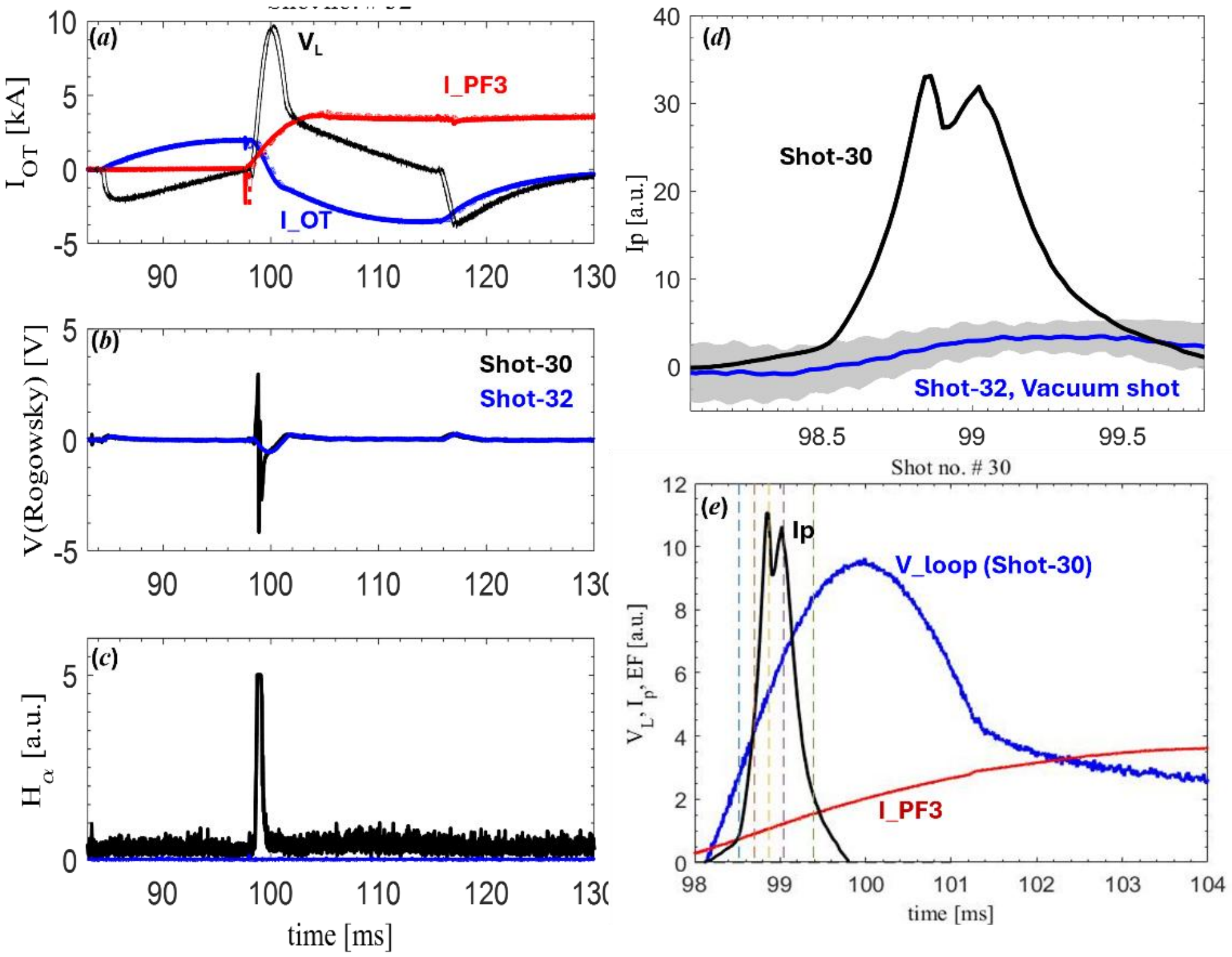

Figure-13: (a) Comparison of vacuum shot (#32) and plasma shot (#30) parameters showing loop voltage, PF3 coil current ($B_z$), and Ohmic coil current, (b) Resulting signals picked up by internal Rogowsky coils for both shots, (c) $H_\alpha$ emission, (d) Integrated rogowsky signal showing plasma current formation in shot #30 and the corresponding background signal in vacuum shot. The shaded portion are the 2σ spread of similar vacuum shots, (e) transient plasma current formation during rising phase of the loop voltage and vertical lines corresponding to the time stamps where vacuum magnetic fields are plotted in figure 15.

After the first plasma commissioning of ST device, regular GDC is carried out and experiments are conducted to optimize vertical field by varying PF3 coil current and timing during the loop voltage formation to initiate plasma current. A few preliminary observations suggest a small plasma current is formed when a field null configuration is achieved transiently during increasing phase of the loop voltage as measured by an internal Rogowsky coil [37- Sameer et al]. Figure 13 presents a comparison between plasma shot #30 and vacuum shot #32, with similar experimental conditions. Shot no 30 shows the formation of a transient plasma current under the applied loop voltage for a duration of 1ms. In the vacuum shot the Rogowsky signal remains near the noise floor (fig 13b) and further comparing similar vacuum shots the 2σ envelope remains close to zero (fig 13d) confirming that the observed plasma current signal is not an artifact of coil pickup or electromagnetic interference. The transient nature of the current is further illustrated in the expanded view in fig 13e. Plasma current begins to rise during the ascending phase of the loop voltage, reaches a maximum near 99 ms, and then decays even as $V_L$ continues to increase. It is worth investigating the reason for such observation.

As mentioned in section 3.1, a substantial toroidal vessel current (~ 15kA) was picked up during magnetic field measurement at the outboard part of vacuum vessel. Since the ST vessel is toroidally continuous and subjected to a high $dI_{OT}/dt$, induced vessel current could be generated during the loop voltage formation. A filament and FEM based simulation is carried out to assess the toroidal current distribution at various parts of the vessel and their effect on vacuum magnetic fields during the loop voltage formation is analyzed, as given in next subsection.

**3.4 Induced vessel Current Measurement:**

A simple filament-based model is developed to estimate vessel current and its distribution by considering a large number of concentric filaments for vacuum vessel and all active magnetic coils like OT, PF3 and passive coils like PF1, 2. By solving coupled circuit equations taking measured coil currents, induced vessel currents are estimated. It is estimated that 100 kA current flows in the vessel inner wall at HFS, 31 kA in top and bottom plates, while outer wall having a current of 13kA during the peak loop voltage formation (figure 14). In addition to this, a FEM based COMSOL model for ST is also developed to further validate the simple filament model. Details of the model and its validation are reported separately [38-Sheetal et. al.]. Efforts are made to measure total vessel current using a fiber optic based current measurement sensor [39, 40] that is installed in tight space in the inner bore area and a preliminary measurement agrees well with the estimated current time profile. However, the magnitude of current shows disagreement by 20% of which detailed calibration and investigation is undergoing and shall be reported separately.

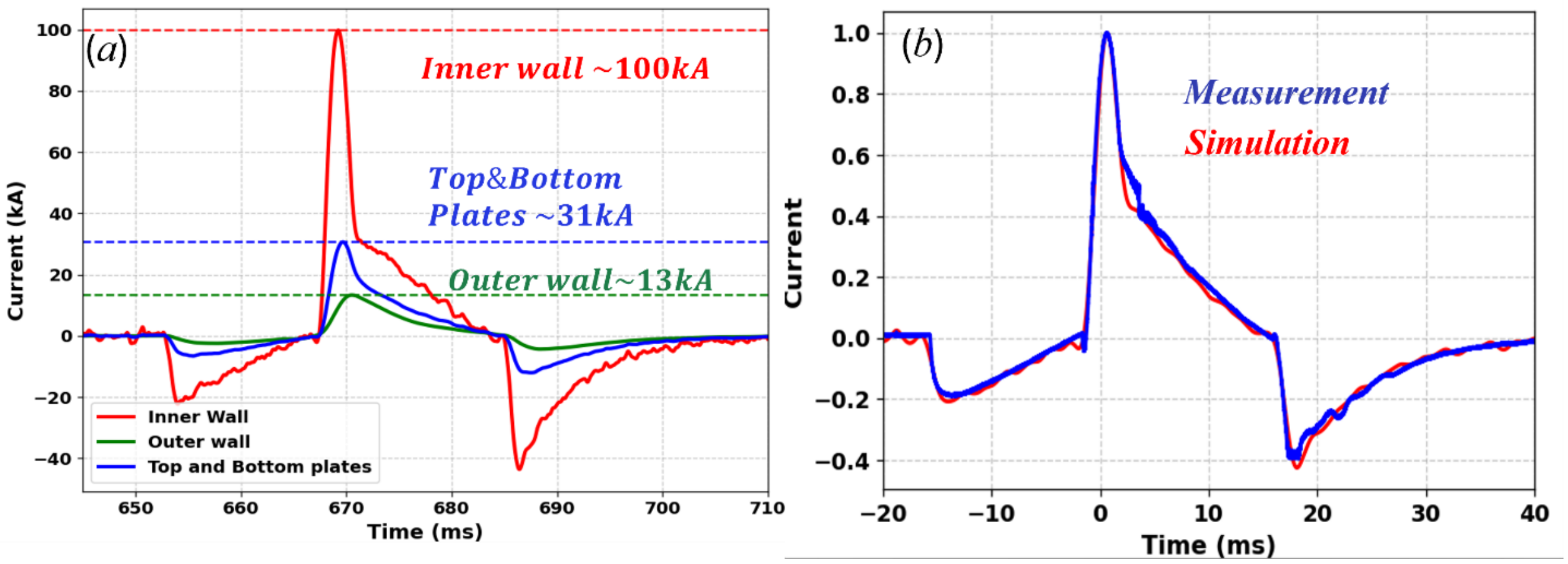

Figure-14: (a) Estimated toroidal vessel current from the filament model as a function of time flowing at various parts of the vacuum vessel, (b) Normalized total vessel current time profile comparison between measurement by fiber optic current sensor and simulation.

The vacuum magnetic fields evolution for the discharge number #30 as shown in figure 13(d) is carried out at the time slices as shown in the same figure. For the present experimental set-up, a vertical downward magnetic field ($-B_z$) is required to support J X B outward force of the expanding plasma current ring once the current is formed. With the available PF-3 current and its ramp rate for the present experiment, the upward $B_z$ component due to induced vessel current overcompensates the applied negative $B_z$ by the PF-3 coils during rising phase of the loop voltage from 98.5 ms to 99.5 ms. This can be seen in figure 15, where the required negative $B_z$ gradually decreases until 98.86ms where plasma current peaks after which the vessel current produces a net positive $B_z$ that may be pushing plasma column towards HFS thus not supporting further plasma current increment.

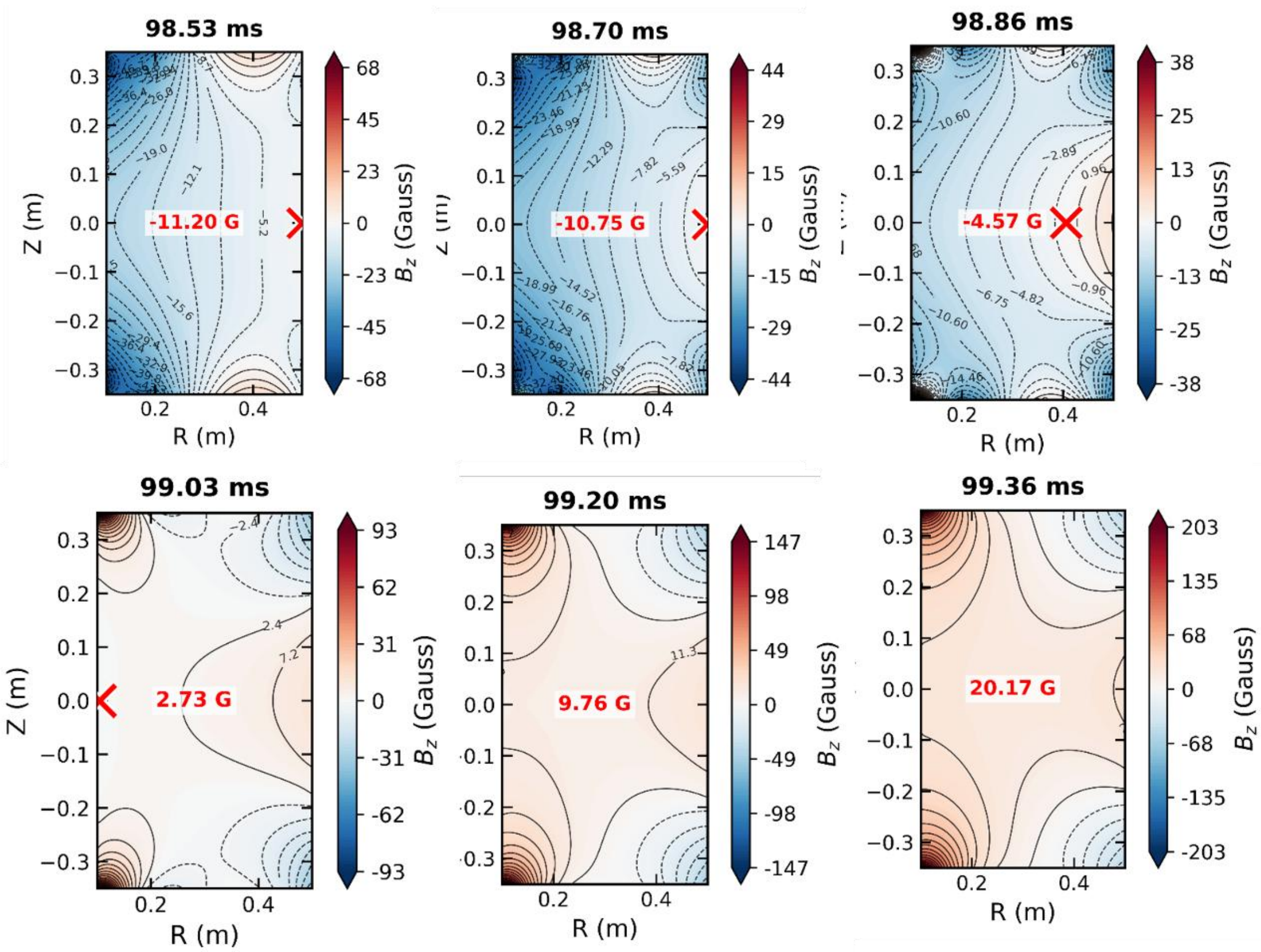

Figure-15: Vacuum magnetic fields computed from the measured OT current, PF-3 coil currents and estimated vacuum vessel current for the shot # 30 at the discrete time slices as sequentially shown in figure 13(e) by broken vertical lines. A supportive negative $B_z$ is established until 98.86 ms post which vessel current dominates to produce a net positive $B_z$ from 99.03 ms onwards.

## 4. Conclusions and Future Prospects:

The new spherical tokamak at IPR is now in operational stage after the system commissioning. All sub-systems have been successfully integrated to the machine and operating satisfactorily. Regular glow discharge conditioning substantially improved the vacuum condition. Further wall conditioning systems are being optimized to study the effect of various DC and RF wall conditioning methods with various gas mixture combinations. Initial plasma experiments (total 114 plasma discharges) indicates transient plasma current formation that appears to be limited by the effect of large induced vessel currents formed during the loop voltage formation. The future work are two folds: 1) Reducing the outboard vessel current 2) Compensating the induced vessel current by suitable optimization of coil configuration or external additional sources. With this background, the ST device termed as India's first spherical tokamak has been successfully commissioned in IPR and plasma experiments have commenced.

**Acknowledgement:**

One of the authors (K.M.) gratefully acknowledges the valuable contributions of Kumarpalsinh Jadeja, Kunal Shah, Rohit Kumar, and J. S. Mishra, as well as members of the ICRH Section (Bhavesh Kadia and Shivam Mishra), and Mitesh Patel from Electronics Group. Authors also thank the various laboratories in the new R&D wing and across IPR for generously providing instruments and accessories during critical stages of the machine integration work. Special thanks are due to the IPR Administration, Purchase, Stores & Accounts Sections for timely executions of requests and IPR Workshop & Drafting Section for their prompt support with urgent fabrication and assembly requirements throughout the integration process.

## References:

1. Y-K.M. Peng and D.J. Strickler 1986 *Nucl. Fusion* 26 769, DOI 10.1088/0029-5515/26/6/005
2. Y.-K. M. Peng, The physics of spherical torus plasmas, Phys. Plasmas 7, 1681–1692 (2000) https://doi.org/10.1063/1.874048
3. A Sykes, R Akers, L Appel, P G Carolan, N J Conway, M Cox, A R Field, D A Gates, S Gee, M Gryaznevich, T C Hender, I Jenkins, R Martin, K Morel, A W Morris, M P S Nightingale, C Ribeiro, D C Robinson, M Tournianski, M Valovic, M J Walsh and C WarrickHide, High-beta performance of the START spherical tokamak, Plasma Phys. Control. Fusion 39 B247, DOI 10.1088/0741-3335/39/12B/019
4. A. Sykes, R. J. Akers1, L. C. Appel1, P. G. Carolan1, J. W. Connor1, N. J. Conway1, G. F. Counsell1, A. Dnestrovskij1,2, Yu. N. Dnestrovskij1,2 et al., H-mode Operation in the START Spherical Tokamak, Phys. Rev. Lett. 84, 495, 2000, DOI: https://doi.org/10.1103/PhysRevLett.84.495
5. A. Sykes et al., "The spherical tokamak programme at Culham," Nuclear Fusion, Vol. 39, No. 9Y, pp. 1271–1281, 1999. DOI: 10.1088/0029-5515/39/9Y/305
6. Kazuaki HANADA et al., Steady-State Operation Scenario and the First Experimental Result on QUEST, Plasma and Fusion Research: Regular Articles Volume 5, S1007 (2010), DOI: 10.1585/pfr.5.S1007
7. A. Sykes, J.-W. Ahn, R. Akers, E. Arends, P. G. Carolan et al., First physics results from the MAST Mega-Amp Spherical Tokamak, Phys. Plasmas 8, 2101 (2001); doi: 10.1063/1.1352595
8. M. Ono et al, Overview of the initial NSTX experimental results, 2001 Nucl. Fusion 41 1435, DOI 10.1088/0029-5515/41/10/311
9. G. D. Garstka; S. J. Diem; R. J. Fonck; B. T. Lewicki; A. C. Sontag; K. L. Tritz; E. A. Unterberg, Performance and stability of near-unity aspect ratio plasmas in the Pegasus Toroidal Experiment, Phys. Plasmas 10, 1705–1711 (2003),https://doi.org/10.1063/1.1559972
10. Morris, AW et al.- MAST Upgrade Divertor Facility: A Test Bed for Novel Divertor Solutions; IEEE transactions on plasma science, Vol 46, issue 5 May 2018:https://doi.org/10.1109/TPS.2018.2815283
11. Menard, JE et al. - Fusion nuclear science facilities and pilot plants based on the spherical tokamak; Nuclear fusion, Vol 56, No 10, 2016: https://iopscience.iop.org/article/10.1088/0029-5515/56/10/106023

12. Adam Baker, The Spherical Tokamak for Energy Production (STEP) in context: UK public sector approach to fusion energy, Phil. Trans. R. Soc. A (2024) 382 (2280): 20230401, https://doi.org/10.1098/rsta.2023.0401
13. Stuart I. Muldrew et al., Conceptual design workflow for the STEP Prototype Powerplant, Fusion Engineering and Design, Volume 201, April 2024, 114238
14. E.T Cheng, Y.K. Martin Peng, Ralph Cerbone, Paul Fogarty, John D Galambos, E.A Mogahed, Brad Nelson, Massoud Simnad, Igor Sviatoslavsky, Mark Tillack, Study of a spherical tokamak based volumetric neutron source, Fusion Engineering and Design, Volume 38, Issue 3, 1998, Pages 219-255, https://doi.org/10.1016/S0920-3796(97)00096-3.
15. T.C. Hender, G.M. Voss, N.P. Taylor, Spherical tokamak volume neutron source, Fusion Engineering and Design, Volume 45, Issue 3, 1999, Pages 265-279, https://doi.org/10.1016/S0920-3796(99)00069-1.
16. F. Troyon And R. Gruber , A Semi-Empirical Scaling Law For The B-Limit In Tokamaks, Physics Letters-A Volume 1 , number 1, 1985
17. F Troyon et al, MHD-Limits to Plasma Confinement, 1984 Plasma Phys. Control. Fusion 26 209
18. J.E. Menard et al,Overview of NSTX Upgrade initial results and modelling highlights, 2017 Nucl. Fusion 57 102006, DOI 10.1088/1741-4326/aa60,
19. JR Harrison et. al., Overview of new MAST physics in anticipation of first results from MAST Upgrade, Nuclear Fusion, 2019
20. Hanada et. al.,Overview of recent progress on steady state operation of all-metal plasma facing wall device QUEST, Nuclear Materials and Energy, Volume 27, June 2021, 101013, https://doi.org/10.1016/j.nme.2021.101013
21. R. Majeski et al., “The Lithium Tokamak Experiment (LTX),” Nuclear Fusion, Vol. 49, No. 5, 055014, 2009. DOI: 10.1088/0029-5515/49/5/055014
22. S Banerjee, DP Boyle, A Maan, N Ferraro, G Wilkie, R Majeski, M Podesta, et. al., Investigating the role of edge neutrals in exciting tearing mode activity and achieving flat temperature profiles in LTX-β, Nuclear Fusion 64 (4), 046026
23. Bhattacharyay et. al.,Study on wall recycling behaviour in CPD spherical tokamak, Fusion Engineering and Design Volume 83, Issues 7–9, December 2008, Pages 1114-1119, https://doi.org/10.1016/j.fusengdes.2008.05.009
24. H. Zushi et al., “Active particle control experiments and critical particle flux discriminating between the wall pumping and fuelling in the compact plasma wall interaction device CPD spherical tokamak,” Nuclear Fusion, Vol. 49, No. 5, 055012 (or nearby), 2009. DOI: 10.1088/0029-5515/49/5/055020
25. S.K. Sharma et. al., Analysis of the footprint traces on the first walls of the compact plasma wall interaction device (CPD) using surface analysis and electron orbit calculations, 2010 Nucl. Fusion 50 025017, DOI 10.1088/0029-5515/50/2/025017
26. Yoshinaga T. et al 2008 Non Inductive formation of spherical tokamak plasmas by ECH on CPD 14th Int. Congress on Plasma Physics (Fukuoka, Japan).
27. Bhatt, S. B., Bora, D., Buch, B. N., & and others, . (1989). Aditya : the first Indian tokamak. Indian Journal of Pure and Applied Physics, 27(9-10), 710–742.

28. R.L. Tanna et al, Overview of physics results from the ADITYA-U tokamak and future experiments, 2024 Nucl. Fusion 64 112011, DOI 10.1088/1741-4326/ad3c50
29. S.P. Deshpande, "SST-1: an overview", 17th IEEE/NPSS Symposium Fusion Engineering (Cat. No.97CH36131), vol.1, pp.227-232 vol.1, 1997.
30. Saxena, Y. C., & SST-1 Team. (2005). First experiments with SST-1 tokamak (Number no. 25/CD, p. [8 p.]).https://inis.iaea.org/records/012v2-b1m23
31. H Idei et. al., Fully non-inductive second harmonic electron cyclotron plasma ramp-up in the QUEST spherical tokamak, Nuclear Fusion 57 (12), 126045
32. K. Mishra et al., “Self organization of high βp plasma equilibrium with an inboard poloidal magnetic field null in QUEST,” Nuclear Fusion, Vol. 55, No. 8, 083009, 2015. DOI: 10.1088/0029-5515/55/8/083009
33. Ranjithkumar et. al., Engineering design aspects of India’s First Spherical Tokamak. Communicated 2026
34. Urmil Thaker, Vaibhav Ranjan, Supriya A. Nair, Development of prototype power supply for ohmic transformer system of SSST, Fusion Engineering and Design, Volume 196, November 2023, 114016
35. Urmil Thaker, Suryakant B. Gupta, Devansh Desai , Ayush, Ankit, Chandan Danani, Kishore Mishra, Supriya Nair, Anil Bharadwaj, Investigation, Conceptual Design, Analysis and Preliminary Test Results of Toroidal Field Power Supply for First Indian Spherical Tokamak: Under review 2026
36. Shekar Goud Thatipamula et. al., Magnetic field measurements and error field estimation in a spherical tokamak: Communicated 2026
37. Sameer et. al., Magnetic Diagnostics of India's First Spherical Tokamak. Communicated 2026
38. Sheetal et. al., Modeling of Induced Vessel Current in the Vacuum Vessel of India’s First Spherical Tokamak at IPR: Communicated 2026
39. Pandya, S.P., Assudani, K., V., P.E., Lachwani, L.T., Jha, S.K., Gopalakrishna, M.V., Pathak, S.K., 2021. Initial Lab Test Results of Magneto Optic Current Sensor Diagnostic Developed for Plasma Current Measurement in Tokamaks. Technical Report IPR/TR-614/2021. Institute for Plasma Research.
40. Pandya, S.P., Tahiliani, K., Suresh, I., V., P.E., Shiroya, A., Lalwani, J., Raval, A., Kathale, H., Jha, S.K., Lachwani, L.T., Aich, S., Kumar, R., Pathak, S.K., Tanna, R.L., Ghosh, J., Daniel, R., Team, A.U.T., 2026. First Results of Magneto-Optic Current Sensor Diagnostic Developed for Plasma Current Measurements in ADITYA-Upgrade Tokamak. Internal Report. Institute for Plasma Research. To be submitted to Fusion Engineering and Design.